\documentclass[ijoc,nonblindrev]{informs3_modified} 

\OneAndAHalfSpacedXII 

\usepackage{natbib}
 \bibpunct[, ]{(}{)}{,}{a}{}{,}%
 \def\bibfont{\small}%
\TheoremsNumberedThrough     

\EquationsNumberedThrough    

\usepackage{algorithm,algpseudocode}
\usepackage{caption,subcaption}

\begin{document}

\RUNTITLE{On Optimization over Tail Distributions}
\TITLE{On Optimization over Tail Distributions}
\ARTICLEAUTHORS{%
\AUTHOR{Cl\'{e}mentine Mottet}
\AFF{Department of Mathematics and Statistics, Boston University, Boston, MA 02115, \EMAIL{cmottet@bu.edu}}
\AUTHOR{Henry Lam}
\AFF{Department of Industrial Engineering and Operations Research, Columbia University, New York City, NY 10027, \EMAIL{henry.lam@columbia.edu}}
} 

\ABSTRACT{We investigate the use of optimization to compute bounds for extremal performance measures. This approach takes a non-parametric viewpoint that aims to alleviate the issue of model misspecification possibly encountered by conventional methods in extreme event analysis. We make two contributions towards solving these formulations, paying especial attention to the arising tail issues. First, we provide a technique in parallel to Choquet's theory, via a combination of integration by parts and change of measures, to transform shape constrained problems (e.g., monotonicity of derivatives) into families of moment problems. Second, we show how a moment problem cast over infinite support can be reformulated into a problem over compact support with an additional slack variable. In the context of optimization over tail distributions, the latter helps resolve the issue of non-convergence of solutions when using algorithms such as generalized linear programming. We further demonstrate the applicability of this result to problems with infinite-value constraints, which can arise in modeling heavy tails.
}%


\maketitle

%


\section{Introduction}
This paper investigates optimization problems in the form
\begin{equation}
\underset{F}{\sup}\ E_F[h(X)]\text{\ \ subject to\ \ }F\in\mathcal U\label{RO}
\end{equation}
where $E_F[\cdot]$ is the expectation over the random variable $X\in\mathbb R$ distributed under $F$. The feasible region $\mathcal U$ on $F$ encodes information about $X$. We are especially interested in $X$ that has unbounded support, with the function $h(\cdot)$ depending heavily on the tail of $X$.

The motivation in studying \eqref{RO} is to apply it to bound performance measures arising in extreme event analysis. The latter has profound importance in risk management in business, engineering, environmental sciences and other disciplines. In conventional studies, extreme event analysis entails understanding the tail behaviors of random variables or data. A common approach is to use extreme value theory (EVT) that asymptotically justifies fitting data with certain parametric distributions. For instance, the Fisher-Tippett-Gnedenko theorem implies that, under proper regularity conditions (i.e., maximum domain of attraction) and normalization, the maximum of an iid sample converges in distribution to a general extreme value distribution (GEV) \cite{fisher1928limiting,gnedenko1943distribution}. The Pickands-Balkema-de Haan Theorem states that the excess losses (i.e., the overshoots of data above a large threshold) converge to a generalized Pareto distribution (GPD)  \cite{balkema1974residual,pickands1975statistical} as the threshold grows, which leads to the well-known peak-over-threshold (POT) method \cite{leadbetter1991basis}. Many approaches have been suggested to estimate tail parameters based on these results (e.g. \cite{smith1985maximum,hosking1987parameter,hill1975simple,davis1984tail,davison1990models}), as well as other generalizations; see, e.g.,\cite{embrechts2013modelling,embrechts2005quantitative} for some excellent reviews of these methods.


In this paper, we study \eqref{RO} to devise an alternate approach to these conventional methods in estimating extremal quantities. Formulation \eqref{RO} is reasoned from the perspective of robust optimization \cite{ben2009robust,bertsimas2011theory} and unlike EVT, this approach can be nonparametric. The set $\mathcal U$, which is known as the uncertainty or the ambiguity set, can incorporate for instance shape information like monotonicity and convexity (prior belief) and moment-type information (estimated from tail data). When $\mathcal U$ is chosen in a statistically correct manner, the optimal value of \eqref{RO} will give a confidence upper bound on the true quantity of interest $E[h(X)]$. Similar claims hold for the lower bound when the maximization is replaced by a minimization. This alternate approach to EVT is motivated from some documented challenges in using asymptotic approximation due to the scarcity of tail data. For example, the POT method needs to choose a threshold to define the ``tail" portion of data. When the threshold is too small, there exists bias in using the GPD fit; on the other hand, when the threshold is large, there can be a lack of data in the tail portion which leads to a high variance estimate. Thus simultaneously minimizing bias and variance can be difficult in some cases and, moreover, can be very sensitive to the underlying distribution (e.g., \cite{embrechts2013modelling}, p.193). On the other hand, when calibrating $\mathcal U$ in \eqref{RO}, one can, in some sense, afford to choose a smaller cutoff threshold to improve the overall bias-variance trade-off, by putting mild prior (but nonparametric) information. This discussion can be found in \cite{lam2015tail}, and we will elaborate further in Section \ref{sec:overview}.

In this paper, we make two contributions towards solving \eqref{RO}, paying especial attention to the tail issues of $X$. First, if $\mathcal U$ contains monotonicity-based constraints, we provide a technique via integration by parts and change of measures to transform the optimization problem into one that only contains moment constraints. The use of integration by parts is in parallel to so-called Choquet's theory (e.g.,  \cite{popescu2005semidefinite}, \cite{van2015distributionally}) that expresses convex classes of probability distributions as mixtures of their extreme points, but more elementary and is generalized from the work of \cite{lam2015tail} that focuses only on tail convexity. Our development using change of measures shows, in addition, that in general there can be more than one equivalent moment problems in the considered context, where some of them can be computationally more advantageous than the others and thus allows more flexibility when using numerical solvers. Second, we provide a methodology to reformulate a moment problem over infinite support into one over compact supports, by paying the price of an additional slack variable. In the situation of infinite support, a moment-constrained optimization problem may not possess an optimal measure, and techniques such as generalized linear programming (or cutting-plane procedure) may fail to converge, in particular because of ``masses" that escape to infinity. The reduction of such problems into ones with compact support ensures the existence of an optimal solution and thus resolves the potential numerical issues. This result is particularly relevant in our use of \eqref{RO} since the $h(\cdot)$ we consider depends on the infinite-support tail of $X$. We also demonstrate how even infinite-value moment constraints can be handled under this framework; these constraints arise when one imposes $X$ to be heavy-tailed (e.g., Pareto).

We conclude this introduction with a brief review of related literature in robust optimization (RO). Pioneered by \cite{ben1998robust,el1998robust}, it considers decision-making when some parameters in the constraints or objectives are uncertain or nosily observed. It aims to obtain solutions that optimizes the worst-case scenario, among all possibilities of the parameter values within a so-called uncertainty set or ambiguity set (e.g., \cite{bertsimas2011theory}). The formulation studied in this paper is closely related to what is known as distributionally robust optimization (DRO), where the uncertain parameter refers to a probability distribution (e.g., \cite{delage2010distributionally,wiesemann2014distributionally}). In extreme event analysis, such an approach has been used in finding worst-case bounds for copula models \cite{embrechts2006bounds,puccetti2013sharp,wang2011complete,dhara2017worst}. Numerical procedures (e.g., the rearrangement algorithm \cite{puccetti2012computation,embrechts2013model}) have also been studied. In the DRO literature, \cite{delage2010distributionally,goh2010distributionally,bertsimas2005optimal} study moment constraints such as mean and second moments. \cite{bertsimas2013data} studies constraints motivated from test statistics such as Kolmogorov-Smirnov tests and $\chi^2$-test. \cite{hanasusanto2015ambiguous,popescu2005semidefinite,van2015distributionally,li2016ambiguous} study the use of unimodal shape information. Our work is closely related to \cite{lam2015tail} that considers convexity constraints and utilizes an integration by parts technique which we generalize. Other types of constraints include the use of statistical distances such as $\phi$-divergences \cite{ben2013robust} and the Wasserstein metric \cite{esfahani2015data}. In particular, \cite{dey2012incorporating,glasserman2014robust,atar2015robust} use Renyi divergence to capture uncertainty in heavy-tail models, and \cite{blanchet2016distributionally} studies the worst-case behavior among distributions within a neighborhood surrounding a GEV model. Finally, \cite{bandi2015robust} studies robust bounds for systems that are potentially driven by heavy-tailed variates, under a deterministic RO framework.

The remainder of the paper is as follows. Section \ref{sec:overview} overviews our formulation and the statistical implications. Section \ref{sec:solution} presents our results on solving the formulation, including the transformation into families of moment problems, removal of redundant constraints and reformulation into problems with compactly supported domains. Section \ref{sec : Numerical Results} shows a numerical example to illustrate our resulting procedure. Section \ref{sec:conclusion} concludes the paper. The Appendix documents all our technical proofs and an auxiliary algorithm.

\section{Motivation of the Formulation and Statistical Implications}\label{sec:overview}
We start with a target extremal quantity in the form $E[h(X)]$ where $h:\mathbb R\to\mathbb R$ is measurable and $X\in\mathbb R$ is a random variable with distribution function $F$. The $h$ function corresponds to the decision-making task at hand and is stated by the user. Elementary examples include level-crossing probabilities where $h(x)=I(x\geq c)$, excess-mean where $h(x)=(x-c)I(x\geq c)$, or entropic type measure where $h(x)=e^{-\theta x}I(x\geq c)$, for some large $c$.


In estimating $E[h(X)]$, we split the data into a portion that is above a chosen threshold  $a$ (the ``tail" portion) and below $a$ (the ``non-tail" portion). To facilitate discussion, we assume further that $h(x)=0$ for $x<a$, so that $h(\cdot)$ only depends on the tail portion. This assumption is justified from our focus on extremal quantities; on the other hand, if it is not satisfied, one can always separate the estimation problem into $E[h(X);X<a]$ and $E[h(X);X\geq a]$, where the former can be handled using standard statistical tools (e.g., empirical estimation).

Next we represent the statistical uncertainty of the tail portion via constraints. We consider the following three types. In our exposition, we denote $\mathbb N=\{0,1,2,\ldots\}$, $\mathbb R^+$ as the non-negative real line, $F_+^{(j)}$ and $F^{(j)}$ as the $j$-th order right derivative and  the $j$-th order derivative of a function $F$ respectively (assuming they exists).
\\

\noindent\emph{Moments: }Consider $F$ such that
\begin{equation}
E_F[g_j(X)]\leq\gamma_{j,1},\ j\in\mathcal J_1\label{moment type}
\end{equation}
where $\gamma_{j,1} \in \mathbb R$, $g_j:[a,\infty)\to\mathbb R$ are some user-specified moment functions, such as $g_j(x)=I(x\geq a)$, $g_j(x)=(x-a)I(x\geq a)$ or $g_j(x)=(x-a)^2I(x\geq a)$, and $\mathcal J_1$ is a finite index set. Obviously the constraints \eqref{moment type} include equality as well as lower bounds by suitably defining the $g_j$ functions.

In our framework we will enforce the constraints $\underline\gamma_{0,1}\leq \bar F(a)=E_F[I(X\geq a)]\leq\overline\gamma_{0,1}$, where $0\leq \underline\gamma_{0,1}\leq \overline\gamma_{0,1} \leq 1$, so that $g_j(x)=I(x\geq a)$ and $g_j(x)=-I(x\geq a)$ are always included.


For modeling tail, using moment constraints alone can be very conservative. Classical results in moment problems stipulate that the  worst-case distribution subject to only moment constraints are typically finitely supported, with the number of support points bounded by the number of constraints \cite{winkler1988extreme}. This does not capture the shape of tail distributions reasonably encountered in practice. In many cases, one may be able to safely conjecture that the tail has decreasing density, yet this information is not captured with moment conditions.
\\

\noindent\emph{Monotonicity: }To incorporate shape information, we consider including assumptions on the monotonicity of $F$ or its derivatives. In the literature, this is known as monotonicity of order $D$ (e.g., \cite{pestana2001higher,van2015distributionally}). Denote $\mathcal P^D[a,\infty)$ as the set of all probability distribution functions that are $D-1$ times differentiable, and the $D$-th order right derivative exists and is finite and monotone on $[a,\infty)$.  We impose
\begin{equation}
F \in \mathcal P^D[a,\infty)\label{monotonicity type}
\end{equation}
where $D\in\mathbb N$ is user-specified. Obviously, \eqref{monotonicity type} holds with $D=0$ minimally by the definition of $F$. Assumption \eqref{monotonicity type} contains several implicit information, as shown by:
\begin{lemma} \label{lemma : property set PL_new}
For any $D \in \mathbb N\setminus\{0\}$, we have:
\begin{enumerate}
\item \label{lemma: item derivative to 0}
For any $F \in \mathcal P^D[a,\infty)$, $\lim_{x\to\infty} F_+^{(D)}(x) = 0$.
\item The set of distribution functions $\mathcal P^D[a,\infty)$ is non-increasing with respect to $D$ in the sense:$$\mathcal P^D[a,\infty)\subset  \mathcal P^{D-1}[a,\infty) \subset \ldots \subset \mathcal P^0[a,\infty)$$
\item For any $F \in \mathcal P^D[a,\infty)$ and $j \in\left\lbrace 1,\ldots, D\right\rbrace$,  the function $(-1)^{j+1} F_+^{(j)}$  exists and is non-negative and  non-increasing. \label{lemma: item alterning monotonicity}
\end{enumerate}
\label{lemma: item non-decreasing set}
\end{lemma}

Lemma \ref{lemma : property set PL_new} is a small variation of the remarks made in \cite{pestana2001higher} p.~320; we present this variation (and provide the proof in Appendix \ref{sec : technical proofs}) since it is needed for our subsequent discussion. Part  1 of Lemma \ref{lemma : property set PL_new} characterizes the asymptotic behavior of the $D^{th}$-order right derivative to converge to $0$ as $x\to\infty$.  Part 2 further stipulates that monotonicity of a derivative in the distribution tail  implies tail monotonicity of any of its derivatives that are of lower order. To conclude, Part 3 specifies that the direction of monotonicity happens in an ``alternating" manner with respect to the order of derivatives. For $D=1$, $F_+^{(1)}$ can only be non-increasing (instead of non-decreasing) on $[a,\infty)$. This means the tail density exists  and is non-increasing. For $D=2$, $F_+^{(2)}$ can only be non-decreasing (instead of non-increasing). The tail density is then convex instead of concave (the latter can be easily seen to be impossible for a tail density). And so forth for higher $D$.


The main purpose of Assumption \eqref{monotonicity type} is to reduce the conservativeness in capturing tail information using only moment constraints. In practice, however, one would likely be able to visually check the assumption up to at most $D=2$, which can be done by, e.g., assessing the pattern of a density or density derivative estimate.
\\

\noindent\emph{Distributional information at the cutoff threshold: }One can also impose bounds on the derivatives of $F$ at the cutoff threshold
\begin{equation}
\underline\gamma_{j,2}\leq (-1)^{j+1}F_+^{(j)}(a) \leq\overline\gamma_{j,2},\ j \in \mathcal J_2\label{derivative type}
\end{equation}
where $0 \leq\underline \gamma_{j,2} \leq \overline \gamma_{j,2}  < \infty$ and $\mathcal J_2$ is a finite set of positive integers. In fact, we will choose $\mathcal J_2$ to be the empty set if $D =0$ in \eqref{monotonicity type} and a finite subset of $\left\lbrace 1, \ldots, D\right\rbrace$ otherwise. This is because without monotonicity conditions on $F_+^{(j)}$, bounding their respective values at $x=a$ has no direct effect on the distributional behavior beyond $a$. Note that one may also incorporate bounds on positions other than $a$, but we leave out this option in the current work.
\\

\noindent\emph{Optimization formulation: }Putting together \eqref{moment type}, \eqref{monotonicity type} and \eqref{derivative type}, we consider the general formulation
\begin{equation}
\begin{array}{lll}
\underset{F}{\sup} &E_F[h(X)]\\
\text{subject to\ \ }& E_F[g_{j}(X)]\leq \gamma_{j,1} &\text{for all $j \in \mathcal J_1$}\\
& \underline\gamma_{j,2}\leq  (-1)^{j+1} F_+^{(j)}(a) \leq\overline\gamma_{j,2}  &\text{for all $j \in \mathcal J_2$}\\
&F \in \mathcal P^D[a,\infty)\\
\end{array}\label{opt : original inf const}
\end{equation}

Using existing terminology, we say that \eqref{opt : original inf const} is consistent if it has a feasible solution, and solvable if it has an optimal solution. In either case, the optimal objective value of \eqref{opt : original inf const} takes values in $\mathbb R \cup \{ +\infty\}$. When \eqref{opt : original inf const} is inconsistent, we set its optimal value as $-\infty$.\\

We next present an immediate statistical implication in using \eqref{opt : original inf const}. Let $F_{true}$ be the true distribution generating a data set (in the frequentist sense). We have:
\begin{theorem}
Suppose that $\gamma_{j,1}$, $\underline\gamma_{j,2}$ and $\overline\gamma_{j,2}$ are calibrated from data such that
\begin{equation}
P_{data}\left(E_{F_{true}}[g_{j}(X)]\leq \gamma_{j,1} , j \in \mathcal J_1\text{\ \ and\ \ }\underline\gamma_{j,2}\leq  (-1)^{j+1} F_{true}^{(j)}(a) \leq\overline\gamma_{j,2}, j \in \mathcal J_2\right)\geq1-\alpha\label{feasibility}
\end{equation}
where $P_{data}$ denotes the probability generated from the data. Then, if $F_{true}\in\mathcal P^D[a,\infty)$, we have
$$P_{data}(E_{F_{true}}[h(X)]\leq Z^*)\geq1-\alpha$$
where $Z^*$ is the optimal value of optimization problem \eqref{opt : original inf const}.\label{guarantee}
\end{theorem}

\proof{Proof of Theorem \ref{guarantee}:}
If $F_{true}$ lies in the feasible region of \eqref{opt : original inf const}, then $E_{F_{true}}[h(X)]\leq Z^*$ by the definition of $Z^*$. Hence under the assumption $F_{true}\in\mathcal P^D[a,\infty)$, we have \begin{equation}
P_{data}\left(E_{F_{true}}[g_{j}(X)]\leq \gamma_{j,1} , j \in \mathcal J_1\text{\ \ and\ \ }\underline\gamma_{j,2}\leq  (-1)^{j+1} F_{true}^{(j)}(a) \leq\overline\gamma_{j,2}, j \in \mathcal J_2\right)\leq P_{data}(E_{F_{true}}[h(X)]\leq Z^*)
\end{equation}
which concludes the theorem.\hfill \Halmos
\endproof

Theorem \ref{guarantee} is a direct application of the statistical argument in DRO (see, e.g., \cite{bertsimas2014robust}). It suggests to calibrate $\gamma_{j,1},\underline\gamma_{j,2},\overline\gamma_{j,2}$ such that \eqref{feasibility} holds. In the rest of this section, we discuss some examples to motivate our investigation in Section \ref{sec:solution}:\\

\begin{example}[Monotonic tail estimation]
\emph{
Consider estimating the tail interval probability $P(\underline b\leq X\leq\overline b)$ where $\underline b<\overline b$ are some large numbers. Choose a cutoff threshold $a$ that separates the tail portion of the data, with $a<\underline b$. We impose the assumption that the tail density exists and is non-increasing on $[a,\infty)$. This can be assessed, for instance, by plotting the density estimate (see Section \ref{sec : Numerical Results}). Find the $95\%$ normal confidence interval for $\bar F(a)$, given by $[\underline\gamma_{0,1},\overline\gamma_{0,1}]$ where $0\leq\underline\gamma_{0,1}\leq\overline\gamma_{0,1}\leq1$. Then, Theorem \ref{guarantee} implies that the optimization
\begin{equation}
\begin{array}{lll}
\underset{F}{\sup} &P_F(\underline b\leq X\leq\overline b) \\
&\underline\gamma_{0,1}\leq\bar F(a)\leq\overline\gamma_{0,1}\\
&F \in \mathcal P^1[a,\infty)\\
\end{array}
\label{eq:example insurance}
\end{equation}
provides a $95\%$ confidence upper bound for the true $P(\underline b\leq X\leq\overline b)$, under the assumption that the density of $X$ is non-increasing on $[a,\infty)$. Optimization \eqref{eq:example insurance} can be seen to bear a simple solution, given by assigning the maximally allowed probability mass on $[a,\infty)$, which is $\overline\gamma_{0,1}$, uniformly over the range $[a,\overline b]$. This gives an optimal value $\overline\gamma_{0,1}(\overline b-\underline b)/(\overline b-a)$.}\label{example monotonicity}\\
\end{example}

\begin{example}[Monotonic tail estimation with density estimate]
\emph{
Continue with Example \ref{example monotonicity}, this time adding an estimate for the density (assumed to exist) at $a$, namely $f(a) = F^{(1)}_+(a)$. This involves bootstrapping the kernel estimate at $a$ to obtain a 95\% confidence interval for $f(a)$. Suppose one does this jointly with the estimation of $\bar F(a)$ that is Bonferroni-corrected, then the optimization
\begin{equation}
\begin{array}{lll}
\underset{F}{\sup} &P_F(\underline b\leq X\leq\overline b) \\
&\underline\gamma_{0,1}\leq\bar F(a)\leq\overline\gamma_{0,1}\\
&\underline\gamma_{1,2}\leq  F_+^{(1)}(a)\leq\overline\gamma_{1,2}\\
&F \in \mathcal P^1[a,\infty)\\
\end{array}
\label{eq:example insurance1}
\end{equation}
also gives a 95\% confidence upper bound for the true $P(\underline b\leq X\leq\overline b)$. This is under the assumption that the density of $X$ is non-increasing on $[a,\infty)$ and the bootstrap calibration of $\underline\gamma_{1,2},\overline\gamma_{1,2}$ is valid. }

\emph{
The optimal value of \eqref{eq:example insurance1} can be built from that of \eqref{eq:example insurance}. The upper bound on the density $f(a)$ is the maximum possible density of $X$ for $X\geq a$. If the height of the maximally allocated uniform density from the solution of \eqref{eq:example insurance} is within the range $[\underline\gamma_{1,2}, \overline\gamma_{1,2}]$, then this uniform density is optimal for \eqref{eq:example insurance1}.  If the height is larger than $\overline\gamma_{1,2}$, the solution for \eqref{eq:example insurance1} becomes the uniform density that has height $\overline\gamma_{1,2}$ starting from position $a$. The later  holds provided that $\overline\gamma_{1,2}(\overline b - a) \geq \underline \gamma_{0,1}$  and $\overline\gamma_{1,2} \geq \underline \gamma_{0,1}/(\overline b-a)$, which gives an optimal value
\begin{equation}
\min\left\{\frac{\overline\gamma_{0,1}}{\overline b-a},\overline\gamma_{1,2}\right\}(\overline b-\underline b)\label{optimal value example}
\end{equation}
Otherwise, the program is inconsistent.  Note that \eqref{optimal value example} is at most $\overline\gamma_{0,1}(\overline b-\underline b)/(\overline b-a)$, the optimal value of \eqref{eq:example insurance}, if one ignores the Bonferroni adjustment. This illustrates the effect in reducing conservativeness by adding extra constraints. Of course, if one adds too many constraints, then the simultaneous estimation issue can become more prominent.}\label{example density}\\
\end{example}

\begin{example}[Monotonic tail estimation with density and moment information]
\emph{
Continue with Example \ref{example density}. Suppose one makes the further assumption that the first moment of $X$ is finite (which can be assessed by exploratory tools such as the maximum-sum ratio; e.g., \cite{embrechts2013modelling} Chapter 6). One can find the 95\% normal confidence interval for $E[(X-a)_+]$, say given by $[\underline\gamma_{1,1},\overline\gamma_{1,1}]$. Suppose that Bonferroni correction is made. Then
\begin{equation}
\begin{array}{lll}
\underset{F}{\sup} &P_F(\underline b\leq X\leq\overline b) \\
&\underline\gamma_{0,1}\leq\bar F(a)\leq\overline\gamma_{0,1}\\
&\underline\gamma_{1,1}\leq E_F[(X-a)_+]\leq\overline\gamma_{1,1}\\
&\underline\gamma_{1,2}\leq F_+^{(1)}(a)\leq\overline\gamma_{1,2}\\
&F \in \mathcal P^1[a,\infty)\\
\end{array}
\label{eq:example insurance2}
\end{equation}
gives a 95\% confidence upper bound for the true $P(\underline b\leq X\leq\overline b)$. This is under the assumption that the density of $X$ is non-increasing on $[a,\infty)$.}


\emph{
Note that unlike Examples \ref{example monotonicity} and \ref{example density}, the solution for \eqref{eq:example insurance2} is more involved. Section \ref{sec:solution} is devoted to a general solution scheme that includes solving \eqref{eq:example insurance2}.}\label{example moment}\\
\end{example}

\begin{example}[Convex tail estimation]
\emph{
Continue with Example \ref{example moment}. Suppose we now make the additional assumption that the density of $X$ is convex. Hence 
\begin{equation}
\begin{array}{lll}
\underset{F}{\sup} &P_F(\underline b\leq X\leq\overline b) \\
&\underline\gamma_{0,1}\leq\bar F(a)\leq\overline\gamma_{0,1}\\
&\underline\gamma_{1,1}\leq E_F[(X-a)_+]\leq\overline\gamma_{1,1}\\
&\underline\gamma_{1,2}\leq F_+^{(1)}(a)\leq\overline\gamma_{1,2}\\
&F \in \mathcal P^2[a,\infty)\\
\end{array}
\label{eq:example insurance3}
\end{equation}
gives a 95\% confidence upper bound for the true $P(\underline b\leq X\leq\overline b)$, under the assumption that the density of $X$ is convex on $[a,\infty)$.}\\
\end{example}


All the example formulations above do not require parametric assumptions on the data, which aim to alleviate the model bias issue and distinguish from the conventional EVT-based methods. On the other hand, the constructed bounds are potentially conservative as they rely on a worst-case calculation. In fact, the smaller the $a$ one chooses, the more sizable  is the tail portion of the data which typically gives more ``flexibility" to generate a higher optimal value in \eqref{opt : original inf const}. Instead of a bias-variance trade-off in the case of using GPD, our approach has a conservativeness-variance trade-off. Depending on the risk management purpose, one may place correctness a priority over conservativeness or vice versa. The next section presents our results on solving optimization \eqref{RO} which covers all the posited example formulations.

\section{Results on the Properties and Solutions of the Formulation}\label{sec:solution}
We present our main results on the properties and solution structure of optimization problem \eqref{opt : original inf const}. This consists of two parts. Section \ref{sec : SILP} presents the reformulation of \eqref{opt : original inf const} into families of moment-constrained problems. Section \ref{sec : formulation LP}  investigates the reduction of infinite-support moment problems into compact-support ones, including those with infinite-value moment constraints. After these, Section \ref{sec:algo} shows how one can numerically solve the reduced formulation.

\subsection{Reduction to Moment Problems via Integration by Parts and Change of Measures}
\label{sec : SILP}

To start our discussion, we introduce $\mathcal Q(\mathcal C)$ as the collection of all bounded non-negative distribution functions on $\mathcal C\in\mathbb R$, where a distribution function is defined as a function that is non-decreasing and right-continuous on $\mathcal C$ (but not necessarily bounded by 1, as in the case of probability distributions). Note that a distribution function as defined is a Stieltjes function of a bounded measure on $\mathcal C$ equipped with the Borel $\sigma$-algebra (e.g., \cite{durrett2010probability}). Correspondingly, we define $\mathcal Q^D(\mathcal C)$ as the collection of all bounded non-negative distribution functions on $\mathcal C$ that are differentiable up to order $D-1$ and have monotone $D$-th order right derivatives. To avoid ambiguity, any distribution function on $\mathcal C$ is defined to take value 0 on $\mathbb R\setminus\mathcal C$.

We first redefine the decision variables in \eqref{opt : original inf const} to be in the space of $\mathcal Q^D[a,\infty)$:
\begin{lemma}
Suppose $\underline\gamma_{0,1}\leq\bar F(a)\leq\overline\gamma_{0,1}$ is included in the first set of constraints in \eqref{opt : original inf const}, where $0\leq\underline\gamma_{0,1}\leq\overline\gamma_{0,1}\leq1$. Then \eqref{opt : original inf const} can be replaced by
\begin{equation}
\begin{array}{lll}
\underset{F}{\sup} &\int h dF\\
\text{subject to\ \ }& \int g_{j}dF\leq \gamma_{j,1} &\text{for all $j \in \mathcal J_1$}\\
& \underline\gamma_{j,2}\leq  (-1)^{j+1} F_+^{(j)}(a) \leq\overline\gamma_{j,2}  &\text{for all $j \in \mathcal J_2$}\\
&F \in \mathcal Q^D[a,\infty)\\
\end{array}\label{opt : original inf const2}
\end{equation}\label{lemma: new}
\end{lemma}

\proof{Proof of Lemma \ref{lemma: new}:}
The lemma follows immediately by checking that $\underline\gamma_{0,1}\leq\bar F(a)\leq\overline\gamma_{0,1}$ enforces the required properties of probability distributions missing in the definition of $\mathcal Q^D[a,\infty)$.\hfill\Halmos
\endproof


This subsection discusses how Program \eqref{opt : original inf const2} can be reformulated into a generalized moment problem in the form
\begin{equation}
 \begin{array}{lll}
\underset{P}{\sup}&\int H dP \\
\text{subject to\ \ }& \int G_{j} dP \leq \gamma_{j} &\text{\ \ for all $j \in \mathcal J$}\\
& P \in \mathcal Q(\mathbb R^+)
\end{array}\label{opt : SILP}
\end{equation}
where $H:\mathbb R^+ \to \mathbb R$ and $G_j:\mathbb R^+ \to \mathbb R$ are measurable functions, $\gamma_j\in\mathbb R$ and $\mathcal J$ is a finite index set.
Program \eqref{opt : SILP} resembles the classic moment-constrained optimization except that the decision variable represents a measure that does not necessarily add up to one. This familiar form allows the adoption of existing optimization routines, as we will discuss in the sequel. The add-up-to-one constraints could be missing in our reformulation because our available mass, i.e., $\bar F(a)$, could be specified in an interval rather than set to be a constant (e.g., 1).


To precisely describe the $H$ and $G_j$ functions,  we introduce some further definitions. Define the function $u_a$ as the shift operator $u_a(x) = x+a$ for all $x\in \mathbb R^+$. In addition, let  $g^{(-d)}(x)$ be the $d$-th order anti-derivative recursively defined as $g^{(-d)}(x) = \int_0^x g^{(-d+1)}(u)du$ and $g^{(0)}\equiv g$. Note that by definition $g^{(-d)}(0)=0$ for any $d>0$.

We introduce Theorem \ref{theorem : reformulation SILP} which requires the following assumption:

\begin{assumption}\label{assump:h}
In program \eqref{opt : original inf const2}, the functions $h$ and $g_{j,1}$ for $j\in\mathcal J_1$ are locally integrable and are either bounded from above or below.
\end{assumption}

\begin{theorem}[Equivalence with a Family of Moment Problems]\label{theorem : reformulation SILP}
Let $D \in \mathbb N\setminus\{0\}$. Denote $Z^*$ as the optimal value of program \eqref{opt : original inf const2}, with $\mathcal J_2\subset\{1,\ldots,D\}$. Under Assumption \ref{assump:h}, we have

\begin{equation}
 \begin{array}{lll}
Z^* = \underset{P}{\sup}&\int H dP \\
\text{subject to\ \ }& \int G_{j,1}  dP\leq \gamma_{j,1} &\text{\ \ for all $j \in \mathcal J_1$}\\
& \underline \gamma_{j,2}\leq \int G_{j,2}  dP\leq \overline \gamma_{j,2} &\text{\ \ for all $j \in \mathcal J_2$}\\
& P \in \mathcal Q(\mathbb R^+)
\end{array}\label{opt : SILP not condensed}
\end{equation}
where for all $x \in \mathbb R^+$,

\begin{itemize}
\item $H(x) = x^{J-D}(h \circ u_a)^{(-D)}(x) $
\item  $G_{j,1}(x)=  x^{J-D}(g_j \circ u_a)^{(-D)}(x)$, for all $j \in \mathcal J_1$
\item $G_{j,2}(x) =  \displaystyle\frac{x^{J-j} }{(D-j)!} $, for all $j \in \mathcal J_2$
\end{itemize}
for any integer $J\in\{0,\ldots,D\}$ (we suppress the dependence of $H$, $G_{j,1}$ and $G_{j,2}$ on $J$ for convenience). In addition,  if program \eqref{opt : SILP not condensed}  is solvable with solution $P^*$, then $F^*$  defined via 
\begin{equation}\label{eq:optimal F}
 D! F^*(x+a) =  \int u^J\left(1 - (1-x/u)^{D}I(u > x) \right) dP^*(u) \ \  \mbox{for all } x \in \mathbb R
\end{equation}
is an optimal solution of program \eqref{opt : original inf const2}.
\end{theorem}
Theorem \ref{theorem : reformulation SILP} is proved by a sequential application of integration by parts and the use of monotonicity to control the tail decay rate of $F$ and its derivatives, by generalizing the techniques in \cite{lam2015tail}. The flexibility in choosing $J$ comes from a change of measure argument, where the decision variable, i.e., the measure of $P$ can be re-expressed as another measure with a likelihood ratio adjustment. As far as we know, using changes of measure to reformulate moment problems is new in the literature, and offers some benefits as we will explain below. Details of the derivation are shown in Appendix \ref{sec : technical proofs}.

A natural choice of $J$ is to set $J = 0$ if $\mathcal J_2$ is empty and $J = \max\{j \in \mathcal J_2\}$  otherwise. Such a choice of $J$ also ensures that $G_{J,2}(x) = 1/(D-J)!$ for all $x \in \mathbb R^+$, and thus we have the constraint $\underline \gamma_{J,2}(D-J)! \leq \int d P \leq  (D-J)!\overline \gamma_{J,2}$ so that the distribution function $P$  in \eqref{opt : SILP not condensed} has bounded mass. Furthermore, when $\mathcal J_2$ is not empty, the feasible set of \eqref{opt : SILP not condensed} can be further restricted  to $\mathcal P(\mathbb R^+)$. This is because the functions $H$, $G_{j,1}$ and $G_{j,2}$ are by construction equal to 0 at $x=0$, so that we can always add an arbitrary mass at 0 to reach the upper bound $(D-J)!\overline \gamma_{J,2}$. This in turn deduces that upon proper normalization of the measure we can impose the constraint that $\int dP=1$.

Another advantage of the above choice of $J$ is that if $h$ and $g_j$ are polynomials, then the reformulation with such a choice of $J$ will also give rise to polynomial forms for $H$, $G_{j,1}$ and $G_{j,2}$. This class of problems can be more susceptible to specialized solution techniques (e.g., semidefinite programming, though this is not the focus of this paper).


Theorem \ref{theorem : reformulation SILP} also reveals the optimality structure of \eqref{opt : original inf const2} in relation to the derivative-based constraints. It is well-known in the theory of moment problems (under non-negative measures) that it suffices to consider $P$ in \eqref{opt : SILP not condensed} that is piecewise constant, i.e., $P$ corresponds to a finite-support distribution (e.g., \cite{rogosinski1958moments}). Therefore, with \eqref{eq:optimal F}, we deduce that it suffices in \eqref{opt : original inf const2} to consider linear combinations of distributions in the form $(1 - (1-x/u)^{D}I(u > x))u^J$ for some $u$, i.e.,
\begin{equation}
F(x+a)=\sum_{i=1}^{N} p_i x_i^J\left(1 - (1-x/x_i)^{D}I(x_i > x) \right),\text{\ \ for\ }x> 0\label{optimal form}
\end{equation}
where $p_i,x_i\geq0$. In other words, either \eqref{opt : original inf const2} is solvable with a solution in the form \eqref{optimal form}, or there exists a sequence of solutions in the form \eqref{optimal form} whose evaluated objective values in \eqref{opt : original inf const2} converge to the optimal objective value $Z^*$. The number $N$ in \eqref{optimal form} is at most the total number of linearly independent  functions in the set $\left\lbrace \left(G_{j,1}\right)_{j \in \mathcal J_1}, \left(G_{j,2}\right)_{j \in \mathcal J_2}, 1\right\rbrace $. This representation is consistent with the notion of generating sets studied in \cite{popescu2005semidefinite}, where in our case $u^J\left(1 - (1-x/u)^{D}I(u > x) \right)$ can be viewed as a generating set. \cite{popescu2005semidefinite}, however, focuses on constraints on the whole distribution and as such, the weights in that context must be probability weights.


In the special case $\mathcal J_2 = \emptyset$, $\mathcal J_1 = \{0,1,2\}$, and $G_{j,1}(x) = (x-a)_+^{j}$ for all $j \in \mathcal J_1$, by setting $J=0$ in Theorem \ref{theorem : reformulation SILP} we arrive at the result given in Theorem 2.1 in \cite{van2015distributionally} in their considered case of univariate $D$-monotone distributions.
We close this subsection by depicting the specific case where the moments are powers of the overshoot variable:



\begin{corollary}\label{corly : finite LP monomial}  If the first set of constraints in program \eqref{opt : original inf const2} is replaced by
$$\underline  \gamma_{j,1} \leq \int (x-a)_+^{j}  dF(x) \leq   \overline \gamma_{j,1} \text{\ \ for all $j \in \mathcal J_1$}$$
where $-\infty \leq \underline  \gamma_{j,1} \leq \overline \gamma_{j,1} <  \infty$, then the conclusion of Theorem \ref{theorem : reformulation SILP} holds with the first set of constraints in \eqref{opt : SILP not condensed} replaced by
$$\underline  \gamma_{j,1} \leq \int G_{j,1}  dP \leq  \overline \gamma_{j,1} \text{\ \ for all $j \in  \mathcal J_1$}$$
where for all $j\in\mathcal J_1$, $G_{j,1}(x) = \Gamma(j+1)/\Gamma(j+D+1) x^{j+J}$, with $J$ set to be any integer in $\{0,\ldots,D\}$,
and $\Gamma(\cdot)$  is the standard Gamma function.
\end{corollary}

\proof{Proof of Corollary \ref{corly : finite LP monomial}:}
The Corollary trivially follows from Theorem \ref{theorem : reformulation SILP} with  functions $(x-a)^j_+$ and $-(x-a)^j_+$ both put into the set of $g_{j,1}(x)$'s.\hfill\Halmos
\endproof

\subsection{Reduction to Compactly Supported Moment Problems}\label{sec : formulation LP}


Our next result shows how one can reduce the moment problem \eqref{opt : SILP} into one whose measures in consideration (i.e., the decision variable) take domain on a compact set. The reason we pursue such a reduction is its requirement to adopt the generalized linear programming technique \cite{goberna1998linear}, which sequentially looks for new support points and updates the probability distributions (more details in Section \ref{sec:algo}). Note that \eqref{opt : SILP} admits feasible measures on the whole non-negative real line (a consequence that our problem focuses on the tail region). As a result, instead of possessing an optimal measure, there may only exist a sequence of measures, whose values converging to the optimal, that possess masses gradually moving to $\infty$ (i.e., such a sequence of measures does not converge weakly; see, e.g., \cite{lam2015tail}). This violates the sufficiency conditions needed for carrying out the generalized linear programming procedure (Theorem 11.2 in \cite{goberna1998linear}) and may potentially deem the procedure non-convergence. In contrast, under the reformulation with compactly supported feasible measures, there always admits an optimal solution (with a finite number of support points) and such an algorithmic issue can be avoided.


%

We would need strong duality to substantiate our results in this subsection. Assume that

\begin{assumption}\label{slater}
Suppose program \eqref{opt : SILP} has a representation in the form
\begin{equation}
\begin{array}{lll}
\underset{P}{\sup}&\int H dP \\
\text{subject to\ \ }& \int\tilde G_{j} dP \leq\tilde\gamma_{j} &\text{\ \ for all $j \in \tilde{\mathcal J}$}\\
&\int\tilde G_{j} dP=\tilde\gamma_{j} &\text{\ \ for all $j \in \tilde{\mathcal J}'$}\\
& P \in \mathcal Q(\mathbb R^+)
\end{array}\label{strong duality form}
\end{equation}
where $\tilde G_j:\mathbb R^+ \to \mathbb R$ are measurable functions, $\tilde\gamma_j\in\mathbb R$ and $\tilde{\mathcal J},\tilde{\mathcal J}'$ are finite index sets, such that there exists $P\in\mathcal Q(\mathbb R^+)$ with $\int\tilde G_jdP<\tilde\gamma_j$ for all $j\in\tilde{\mathcal J}$, and $(\tilde\gamma_j )_{j\in\tilde{\mathcal J}'}$ is in the interior of the set
$$\left\{\left(\int\tilde G_{j}dP\right)_{j\in\tilde{\mathcal J}'}:P\in\mathcal Q(\mathbb R^+)\right\}$$\end{assumption}

Assumption \ref{slater} is a Slater-type condition for moment problems. Similar assumptions have been documented in, e.g., \cite{bertsimas2005optimal,popescu2005semidefinite,shapiro2001duality,karlin1966tchebycheff,smith1995generalized}. Under Assumption \ref{slater}, strong duality holds for \eqref{opt : SILP}:

\begin{theorem}\label{theorem:strong duality}
Suppose program \eqref{opt : SILP} is consistent and denote $Z^*$ its optimal objective value. If Assumption \ref{slater} is satisfied, then strong duality holds for \eqref{opt : SILP}, $i.e.$ $Z^*$ is equal to
\begin{equation}
\begin{array}{lllll}
\underset{y}{\inf} &\sum_{j \in \mathcal J }  y_j \gamma_j  \\
\text{subject to}&\sum_{j \in \mathcal J} y_j G_j(u) &\geq H(u)&\text{\ \ for all\ }u\in \mathbb R^+\\
&y_j &\geq 0 & \text{\ \ for all\ } j \in \mathcal J
\end{array}\label{opt : dual strong duality}
\end{equation}
\end{theorem}

The proof is a direct application of Lagrangian duality for optimization with both inequality and equality constraints, depicted in Chapter 8, Problem 7  in \cite{luenberger1997optimization}, together with standard weak duality as argued in, e.g., Section 3.1 in \cite{smith1995generalized}.


Next, we exclude some trivial scenarios and redundant constraints. We introduce the collection of index sets $\mathcal J(x)$ defined as
\small\begin{align}
\mathcal J(x) = \left\lbrace i \in  \mathcal J\left|\exists u_n\in\text{supp}(G_i)\text{\ s.t.\ } \limsup_{u_n\to x}\frac{H(u_n)}{\left|G_i(u_n)\right|} \geq 0 \text{\ \ and \ } \limsup_{u_n\to x} \ \frac{G_j(u_n)}{\left|G_i(u_n)\right|}   \leq 0 \text{ for all } j \in  \mathcal J \right.\right\rbrace\label{redundant set}
\end{align}
\normalsize
where $x \in \mathbb R^+ \cup \{\infty\}$ is a placeholder, and $\text{supp}(G_i) $ denotes the support of the function $G_i$, $i.e$ $\text{supp}(G_i) = \{u\in \mathbb R^+| G_i(u) \neq 0\}$.

\begin{theorem}[Removal of Redundant Constraints]\label{theorem:include lambda 0 infty}
Given any fixed $x \in \mathbb R^+ \cup \{\infty \}$. Suppose Assumption \ref{slater} holds. Denote $Z^*$ as the optimal value of \eqref{opt : SILP}. Then the following statements hold:

\begin{enumerate}
\item If $\mathcal J = \mathcal J(x)$ and $\sup\{H(u)| u \in \mathbb R^+\} > 0$,  then $Z^* = \infty$.
\item If $\mathcal J =\mathcal J(x)$ and $\sup\{H(u)| u \in \mathbb R^+\} \leq 0$,  then $Z^* = 0$.
\item  If $\mathcal J(x) \subsetneq \mathcal J$, then
\begin{equation}
\begin{array}{llll}
Z^*=&\underset{P}{\sup}&\int H dP \\
&\text{subject to\ \ }& \int G_{j} dP \leq \gamma_{j} &\text{\ \ for all $j \in \mathcal J \setminus \mathcal J(x)$}\\
&& P \in \mathcal Q(\mathbb R^+)
&\end{array}\label{opt : SILP new}
\end{equation}
\end{enumerate}
\end{theorem}

The set $\mathcal J(x)$ is the set of redundant constraints. Identifying it can screen out the trivial cases (Cases 1 and 2 in Theorem \ref{theorem:include lambda 0 infty}) and reduce the number of constraints (Case 3). We also note that the choice of $x$ in applying Theorem \ref{theorem:include lambda 0 infty} is self-consistent, in the sense that choosing any $x$ gives rise to valid results and, moreover, one can apply the theorem sequentially on different $x$'s.

When $H$ and $G_i$'s ar e continuous at $x$, then the definition of $\mathcal J(x)$ can be reduced to having the inequalities hold for the ratios evaluated at $x$ (by merely considering $u_n=x$). Definition \eqref{redundant set}, however, is more general as it covers discontinuous cases and the case where $x=\infty$.

Appendix \ref{sec : appendix tech proof part 2} provides the proof of Theorem \ref{theorem:include lambda 0 infty}, which relies on analyzing the allowable asymptotic behaviors of solution sequences in relation to the behaviors of $H$ and $G_j$ around $x$.

We have the following simplification in the case where all the constraint functions in \eqref{opt : SILP} have both lower and upper bounds, which can be derived using the definition of $\mathcal J(x)$:
\begin{lemma}\label{rmk:empty J}
Assume \eqref{opt : SILP} can be expressed as
\begin{equation}
\begin{array}{lll}
\underset{P}{\sup}&\int H dP \\
\text{subject to\ \ }& \underline\gamma_j\leq\int\tilde G_{j} dP \leq\overline\gamma_{j} &\text{\ for all\ }j\in\tilde{\mathcal J}\\
& P \in \mathcal Q(\mathbb R^+)
\end{array}\label{formulation empty J}
\end{equation}
for some $\tilde G_j:\mathbb R^+\to\mathbb R$ and finite index set $\tilde{\mathcal J}$, where $-\infty<\underline\gamma_j\leq\overline\gamma_j<\infty$. Then $\mathcal J(x)$ defined in \eqref{redundant set} is empty for all $x \in \mathbb R^+ \cup \{\infty\}$.
\end{lemma}

Our main result in this subsection is to demonstrate how a slack variable $s$ can be introduced to encode the case where some mass ``escapes" to $\infty$. This allows us to reduce the search space of \eqref{opt : SILP new} to compact-support distributions  when some regularity conditions are met.

To this end, when $\mathcal J \setminus \mathcal J(\infty)$ is not empty, we make the  following assumptions.

\begin{assumption}
 For some $M \in \mathcal J \setminus \mathcal J(\infty)$  and  $u$ large enough,  the function $G_M(u)$  is  bounded away from $0$ and $
\limsup_{u \to \infty} \left|G_j(u)/G_M(u)\right| < \infty$ for all $j \in \mathcal J \setminus \mathcal J(\infty)$.
\label{assump:regularityinfty}
\end{assumption}

\begin{assumption} \label{assump:well defined}For some  $M \in \mathcal J \setminus \mathcal J(\infty)$, the limit
 \begin{equation} \lambda_{j,M} =  \underset{u\to \infty}{\lim}\frac{G_j(u)}{\left|G_M(u)\right|}
\end{equation}
is well-defined (on the extended real line) for all $j\in \mathcal J \setminus \mathcal J(\infty)$.
\end{assumption}

Assumptions \ref{assump:regularityinfty} and \ref{assump:well defined} ensure the limits of ratios of $G_j(x)$ and $G_M(x)$ are well-defined as $x\to\infty$, which is needed to handle the situation of escaping mass. Next, we impose some mild regularity conditions on $G_j$ and $H$:

\begin{assumption}\label{assump:semi-cont G}
For all  $j \in \mathcal J \setminus \mathcal J(\infty)$, the functions $G_j$ are lower semi-continuous and bounded on any compact set of $\mathbb R^+$.
\end{assumption}

\begin{assumption}\label{assump:semi-cont H}
The function $H$ is upper semi-continuous and bounded on any compact set of $\mathbb R^+$.
\end{assumption}

Lastly, we assume the following non-degeneracy condition for at least one of the $G_j$'s:

\begin{assumption} \label{assump:postivity}There exists some $j \in \mathcal J \setminus \mathcal J(\infty)$ such that $\inf_{x\in \mathbb R^+} G_j(x) > 0$.
\end{assumption}

Assumption \ref{assump:postivity} can be ensured to satisfy in formulation \eqref{opt : SILP not condensed}  by using $J=\max\{j\in\mathcal J_2\}$ in Theorem \ref{theorem : reformulation SILP} (i.e., through a particular change of measure on the decision variable as in its proof) so that $G_{J,2}=1/(D-J)!$ for all $x\in\mathbb R^+$.

\begin{theorem}[Slack Variable to Encode Escaping Mass]\label{theorem:existence bounded measure}
Suppose \eqref{opt : SILP} is consistent with optimal value $Z^*$. We have:

\begin{enumerate}
\item If $\mathcal J = \mathcal J(\infty)$ and $\sup\{H(u)| u \in \mathbb R^+\} > 0$,  then $Z^* = \infty$.\label{theorem:item empty positive}
\item If $\mathcal J = \mathcal J(\infty)$ and $\sup\{H(u)| u \in \mathbb R^+\} \leq 0$,  then $Z^* = 0$.
\item  If $\mathcal J(\infty) \subsetneq \mathcal J$ and Assumption  \ref{assump:regularityinfty} holds, then
\begin{enumerate}
\item If $\lambda_M :=  \limsup_{u\to \infty} H(u)/|G_M(u)| = \infty $ then $Z^* = \infty$.
\label{theorem:item lambdam inf lambda M inf}
\item Otherwise, if Assumptions \ref{assump:well defined}-\ref{assump:postivity} hold, then there is some $C \in \mathbb R^+$ such that $Z^* < \infty$ and
\begin{equation}
\begin{array}{*{5}l}
Z^* =  \underset{P, s}{\sup}&\int H dP  +  \lambda_M  s  \\
\text{subject to}&  \int G_j dP +  \lambda_{j,M} s   \leq \gamma_j\text{\ \ for all $j \in  \mathcal J \setminus \mathcal J(\infty)$}\\
&s\geq 0\\
& s = 0 \text{ if } \lambda_M  = -\infty \\
& P \in \mathcal Q_{N}[0,C]
\end{array}\label{opt: lambda infty finite support}
\end{equation}
where $N$ is the number of linearly independent functions in the collection  $\left\lbrace (G_j)_{j\in \mathcal J \setminus \mathcal J(\infty)}, 1\right\rbrace$, and $\mathcal Q_{N}[0,C]$ is the set of all distribution functions on $[0,C]$ that are piecewise constant, right-continuous with at most $N$ jumps. In particular, program \eqref{opt: lambda infty finite support} is solvable.
\end{enumerate}
\end{enumerate}
\end{theorem}

Theorem \ref{theorem:existence bounded measure} is proved by tracking the limits of all the possible sequences of weights and support points in a finite-support measure that can tend to the optimal value.

Our next result specializes to handle interval-type power function constraints:

\begin{corollary}\label{corly:moment LP}
Consider Program \eqref{opt : SILP not condensed} where the first set of constraints is replaced by
$$\underline \gamma_{j,1} \leq \int G_{j,1}  dP \leq \overline \gamma_{j,1} \text{\ \ for all $j \in  \mathcal J_1$}$$
$G_{j,1}$ is defined as in Corollary \ref{corly : finite LP monomial}, and $0 \leq \underline \gamma_{j,1}\leq \overline \gamma_{j,1} < \infty$. Set $J $ in Program \eqref{opt : SILP not condensed} to be 0 if $\mathcal J_2$ is empty and $\max\{j \in \mathcal J_2\}$ otherwise. Denote $Z^*$  as the optimal  value, $M = \max\{j \in \mathcal J_1\}$, $\lambda_M = \limsup_{x\to\infty} \ H(x)/|G_{M,1}(x)|$, and  $\delta_{j,M}$ as the Kronecker delta function. In addition, assume that \eqref{opt : SILP not condensed}  is consistent, and that there exists $P \in \mathcal Q(\mathbb R^+)$ such that $\underline \gamma_{j,i} < \int G_{j,i} dP < \overline \gamma_{j,i}$ for all $j$ and $i$ such that $\underline \gamma_{j,i} < \overline \gamma_{j,i}$, and $\left(\left(\underline \gamma_{j,1}\right)_{j\in \tilde{\mathcal J_1}}, \left(\underline \gamma_{j,2}\right)_{j\in \tilde{\mathcal J_2}}\right)$  is an interior point of

\begin{equation}
\left(\left(\left(\int G_{j,1}dQ \right)_{j \in \tilde{\mathcal J_1}}, \left(\int G_{j,2}dQ \right)_{j \in \tilde{\mathcal J_2}}\right):  Q \in \mathcal Q(\mathbb R^+)\right)
\end{equation}
with $\tilde{\mathcal J_i} = \left\lbrace j \in \mathcal J_i| \underline \gamma_{j,i} = \overline \gamma_{j,i}\right\rbrace$ and $i \in \{1,2\}$. Then

\begin{enumerate}
\item If $\lambda_M = \infty$, then $Z^* = \infty$.
\item If $\lambda_M < \infty$ and $h$ is upper semi-continuous when $D = 0$, then $Z^* < \infty$ and there exists some $C \in \mathbb R^+$ such that
\begin{equation}
 \begin{array}{*{5}l}
Z^* =\underset{P, s}{\sup}&\int H dP +  \lambda_{M} s \\
\text{subject to}&  \underline \gamma_{j,1}\leq \int G_{j,1} dP + \delta_{j,M} s \leq \overline \gamma_{j,1} &\text{\ \ for all $j \in \mathcal J_1$}\\
&  \underline \gamma_{j,2}\leq \int G_{j,2} dP  \leq \overline \gamma_{j,2} &\text{\ \ for all $j \in \mathcal J_2$}\\
&s\geq 0\\
& s = 0 \text{ if } \lambda_M  = -\infty\\
& P \in \mathcal Q_{N}[0,C]
\end{array}\label{opt:finite LP moments no J_i empty}
\end{equation}

is solvable and $N = |\mathcal J_1| + |\mathcal J_2| $.
\end{enumerate}
 \end{corollary}

In Theorem \ref{theorem:existence bounded measure} and Corollary \ref{corly:moment LP}, the non-trivial cases of the optimization formulation, namely \eqref{opt: lambda infty finite support} and \eqref{opt:finite LP moments no J_i empty}, are now posited as moment problems with compact support. With this formulation we can apply generalized linear programming (discussed in the next section) without running into the numerical issue of having n

From Theorem \ref{theorem:include lambda 0 infty}, we also derive Corollary \ref{corly:existence P*} below. This last result is of interest as it handles, under some regularity conditions, programs in the form

\begin{equation}
 \begin{array}{lll}
\underset{P}{\sup}&\int H dP\\
 \text{subject to\ \ }& \int G_j dP \leq \gamma_{j} &\text{\ \ for all $j \in \mathcal J$}\\
 &\int |G|dP = \infty \\
 & P \in \mathcal Q(\mathbb R^+)
 \end{array}\label{opt : infinite constraint}
 \end{equation}
where $G:\mathbb R^+ \to \mathbb R$ is a measurable function on $\mathbb R^+$. Program \eqref{opt : infinite constraint} extends the scope of  \eqref{opt : SILP} by including an infinite-value constraint.  Such formulation can arise in practice when the  tail decay is estimated to have infinite moments (e.g., Pareto-type tail, which can be assessed by the ratio-of-maximum-and-sum method; Section 6.2.6 in \cite{embrechts2013modelling}). For example, from Corollary \ref{corly:existence P*}(1) and  \ref{corly:existence P*}(3),  we know that the following optimization with infinite constraint
\begin{equation}
  \begin{array}{lll}
 \underset{P}{\sup}&\int xI(x\geq b)dP(x)\\
 \text{subject to\ \ }& \int x dP(x) \leq \gamma\\
 & \int x^2 dP(x) = \infty\\
 & P \in \mathcal Q(\mathbb R^+)
 \end{array}
 \end{equation}
(for some given $b$) is the same as
\begin{equation}
  \begin{array}{lll}
 \underset{P}{\sup}&\int xI(x\geq b)dP(x)\\
 \text{subject to\ \ }& \int x dP(x) \leq \gamma\\
 & P \in \mathcal Q(\mathbb R^+)
 \end{array}
 \end{equation}

\begin{corollary}[Infinite-value Constraints]\label{corly:existence P*}
Consider Program \eqref{opt : SILP},  denote $Z^*$ as its optimal objective value, and  assume the program is consistent and that Assumption \ref{assump:regularityinfty} holds. In addition, let $G$ be a not identically $0$ and measurable function satisfying $\limsup_{u\to\infty} |G(u)/G_M(u)| = \infty$ when $\mathcal J(\infty) \subsetneq \mathcal J$.  If any one of the following statements holds:
\begin{enumerate}
\item $Z^* = \infty$ and $\liminf_{u\to\infty}|G(u)/H(u)| > 0$
\item $Z^* = \infty$, $\liminf_{u\to\infty}H(u)/|G(u)| > 0$,  and $\inf_{x\in \mathbb R^+} G_j(x) >0$ for some $j \in \mathcal J$
\item $Z^* < \infty$ and $\limsup_{u\to\infty}|G(u)/H(u)| = \infty$
\end{enumerate}
then  \eqref{opt : SILP} and \eqref{opt : infinite constraint} have the same optimal objective value $Z^*$.
 \end{corollary}


\subsection{A Generalized Linear Programming Procedure}\label{sec:algo}

With the results in Sections \ref{sec : SILP} and \ref{sec : formulation LP}, we apply generalized linear programming to solve problems in the form \eqref{opt : SILP}. Our procedure is shown in Algorithm \ref{algo:GKLP}, which is an adaptation of Algorithm 11.4.1 in \cite{goberna1998linear} (setting $\varepsilon_k = 0$ and $|S_k|=1$ for all $k \in \mathbb N$ there). This procedure relies on the sufficiency to search for distributions that have finite support. It iteratively searches for the optimal support points by  looking for the next point that has the highest current ``reduced cost" via solving a ``subproblem" (i.e., the point not already in the set of considered support points that gives the highest rate of improvement by assigning it a mass), and updating the solution via a ``master problem" that is a linear program on the existing support points. Compared to the semidefinite programming approach, the generalized linear programming applies to non-polynomial objective functions and constraints, but in our context it requires solving potentially a non-convex one-dimensional search in each iteration.

Under Theorem \ref{theorem:existence bounded measure}, our formulation is cast over measures with compact support. Theorem 11.2 in \cite{goberna1998linear} guarantees that if the dual optimization is consistent, then Algorithm \ref{algo:GKLP} generates a sequence of dual multipliers $y_j^{(k)}$ that converges to an optimal dual solution and it does so in  a finite number of iterations when $\varepsilon > 0$. When the dual multipliers converge, the  value returned by the procedure is also the optimal objective value of \eqref{opt: lambda infty finite support}, up to some tolerance $\varepsilon$ (Section 11.1 \cite{goberna1998linear}). Note that, in general, a good value of the compact support boundary $C$ is not known a priori. In our experiment, we choose $C$ to be in the tens, which appear to work well.

The initialization step in Algorithm  \ref{algo:GKLP}  can be done by applying a Phase I procedure described in Algorithm \ref{GLP initialization} in Appendix \ref{sec : algo Phase I}, which finds a feasible solution for \eqref{opt: lambda infty finite support} provided such solution exists.  Algorithm \ref{GLP initialization} attempts to solve the following program
\begin{equation}
\begin{array}{llll}
&\underset{ s, r, P}{\min} & r\\
&\text{subject to}& - r + \int G_j(x)dP +  \lambda_{j,M} s\leq \gamma_j ,& \forall j \in \mathcal J\setminus \mathcal J(\infty)\\
&&s, r \geq 0\\
&& s = 0 \text{ if } \lambda_M  = -\infty\\
&&P \in \mathcal Q_N(\mathbb R^+)
\end{array}\label{opt: Phase I}
\end{equation}
If the algorithm stops with $(P^*, s^*,r^*)$ such that $r^* = 0$, $(P^*, s^*)$ is a feasible solution of \eqref{opt: lambda infty finite support}.  If $r^* > 0$, we conclude that \eqref{opt: lambda infty finite support} has no feasible solution. Under conditions in Corollary 4.1 in \cite{magnanti1976generalized}, a variant of this Phase I procedure converges in finite steps (even with tolerance level 0).

\begin{algorithm}[!htp]
\caption{Computing the optimal value of  Program \eqref{opt: lambda infty finite support} when $\mathcal J \setminus \mathcal J(\infty)$ is not empty}
  \label{algo:GKLP}
{\small \textbf{Inputs:} Provide the parameters $\gamma_j$ and the functions $G_j$ for all $j \in \mathcal J\setminus \mathcal J(\infty)$, and the function $H$. Compute the quantities $\lambda_{j,M},j \in \mathcal J\setminus \mathcal J(\infty)$ and $ \lambda_M$. Also specify a big number $C \in  \mathbb R^+$ and a tolerance level $\varepsilon \geq 0$.\\

  \textbf{Exclusion of the trivial scenarios}
  \begin{itemize}
\item IF $\lambda_M$ is equal to $\infty$, then $Z^* = \infty$
\item ELSE proceed to the next step of the procedure
\end{itemize}

  \textbf{Initialization: }
\begin{itemize}
\item Find an initial feasible solution in $\mathcal Q_{L}[0,C]$ for Program \eqref{opt: lambda infty finite support}, where $L \in \{1,\ldots,N\}$ and $N$ is the number of linearly independent functions in the collection $\{(G_j)_{j \in \mathcal J \setminus \mathcal J(\infty)}, 1\}$.  This can be done using the Phase I algorithm in Appendix \ref{sec : algo Phase I}. Denote $(x_i)_{i \in 1\ldots L}$ as the support points of  the initial feasible solution.
\end{itemize}
  \textbf{Procedure: }For each iteration $k=0,1,\ldots$, and given $(x_i)_{i \in 1\ldots L+k}$:

  \begin{algorithmic}
  \State \textbf{1. }(Master problem) Solve
  \begin{equation}
  \label{eq:outer_opt}
  \begin{array}{ll}
Z^k = \underset{p, s}{\sup}&\sum_{i=1}^{L+k} H(x_i)p_i + \lambda_M s \\
\text{subject to}& \sum_{i=1}^{L+k} G_j(x_i)p_i + \lambda_{j,M} s\leq \gamma_j ,\ \ \forall j \in \mathcal J \setminus \mathcal J(\infty)\\
&s \geq 0\\
& s = 0 \text{ if } \lambda_M  = -\infty\\
&p_i\geq 0 \ \ \forall i=1,\ldots,L+k
\end{array}\end{equation}
Let $( p^k, s^k)$ be the optimal solution. Find the dual multipliers  $(y_j^k)_{j \in \mathcal J \setminus \mathcal J(\infty)}$ satisfying
$$\begin{array}{lll}
&\left(\sum_{j \in \mathcal J \setminus \mathcal J(\infty)} y_j^k G_j(x_i^k) - H(x_i^k)\right) p_i^k = 0,\ \forall i=1,\ldots,L+k\\
&\left(\sum_{j \in \mathcal J \setminus \mathcal J(\infty)} y_j^k\lambda_{j,M} -  \lambda_{M}\right)s^k=0\\
&y_j^k \geq 0,  \text{for all $j \in \mathcal J \setminus \mathcal J(\infty)$}\\
\end{array}$$

\State \textbf{2. }(Subproblem) Find $x_{L+k +1}$ that minimizes
\begin{equation}\label{eq:inner_opt}
\rho^k(u) =\sum_{j \in \mathcal J \setminus \mathcal J(\infty)}y_j^{k}G_j(u) - H(u), \ \ \ \text{where } u \in [0,C]
\end{equation}

\begin{itemize}
\item SET $\epsilon^k = \rho^k(x_{L+k+1})$
\item IF $\epsilon^k \geq -\varepsilon$, STOP and RETURN $Z^* = Z^k$ ELSE go back to 1.
\end{itemize}\\

  \end{algorithmic}
  }
  \label{GLP}

\end{algorithm}

\section{Numerical Example}\label{sec : Numerical Results}
\begin{figure}[h]
 \center
 \includegraphics{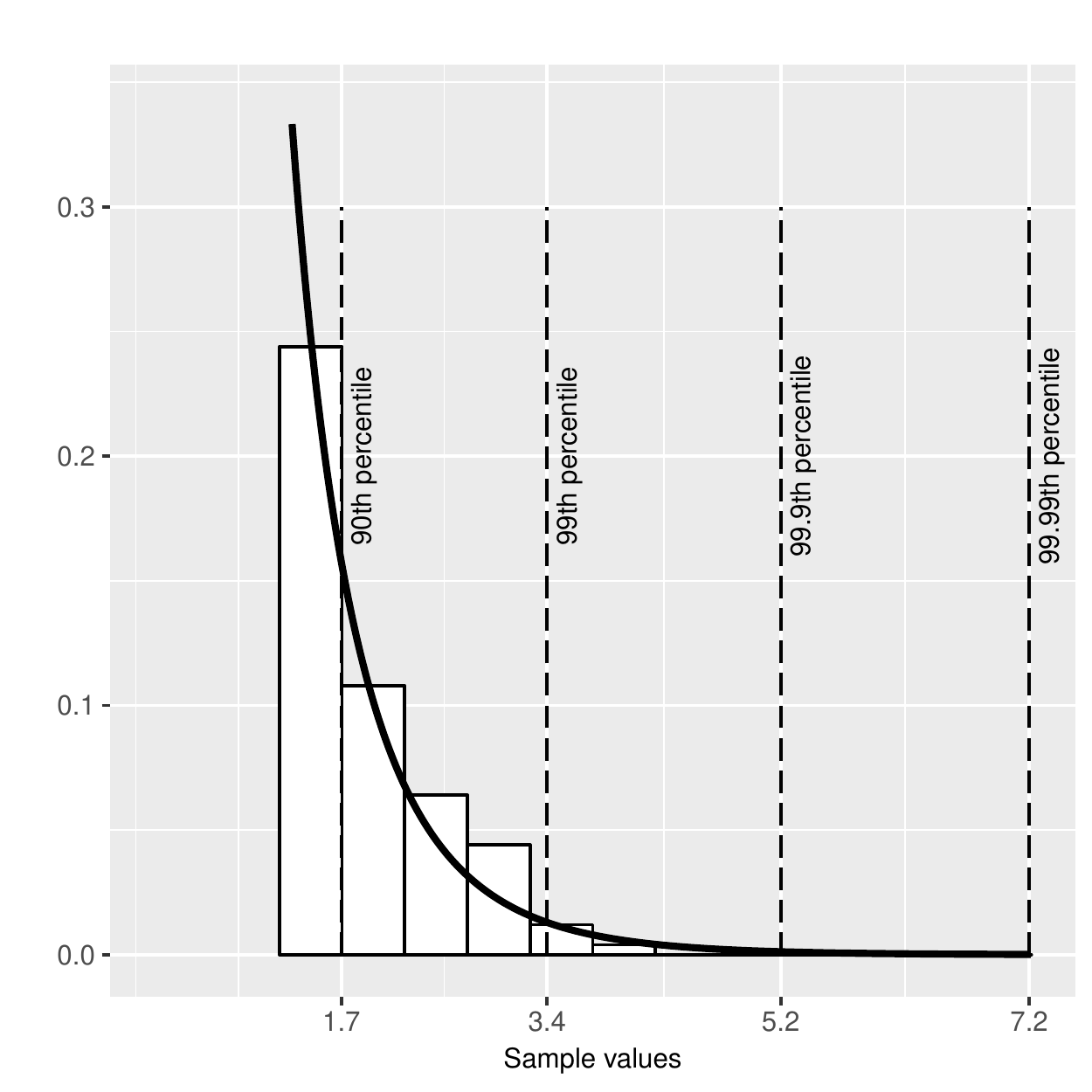}
  \caption{\label{fig : hist Horror} \textit{Normalized histogram of 500 iid observations sampled from the probability distribution function $F_X(x) = 1 - x^{-1}e^{-x}$, where $x \geq x_0$ and $x_0 e^{x_0} = 1$. The thick line is the true probability density function and the dashed lines indicates the theoretical  $p^{th}$-percentiles  when $p \in \{90, 99,99.9,99.99\}$}}
 \end{figure}

We demonstrate our results and procedure in Section \ref{sec:solution} with a numerical example. Figure \ref{fig : hist Horror} is a normalized histogram of 500 observations, each representing an independent realization of the random variable $X$ with distribution function $F_X(x) = 1 - x^{-1}e^{-x}$ for all $x \geq x_0$ where $x_0 e^{x_0} = 1$.  The thick line represents the true probability density function and the dashed lines indicate the  values $q_p$ of the theoretical  $p^{th}$-percentiles associated with $F_X$, when $p \in \{90, 99,99.9,99.99\}$.

We test our procedure in estimating $P(X\geq q_p)$ for the set of $p$'s depicted above. We consider the class of optimization formulations
\begin{equation}
\begin{array}{lll}
\underset{F}{\sup} &P_F(X\geq q)\\
\text{subject to\ \ }&\underline\gamma_{j,1}\leq E_F[(X-a)_+^j]\leq\overline\gamma_{j,1} &\text{for all $j \in \mathcal J_1$}\\
& \underline\gamma_{j,2}\leq (-1)^{j+1}F_+^{(j)}(a) \leq\overline\gamma_{j,2}  &\text{for all $j \in \mathcal J_2$}\\
&F \in \mathcal P^D[a,\infty)\\
\end{array}\label{opt : original inf const numerics}
\end{equation}
where $D $ can be any value in $\{0, \ldots, 5\}$, $\mathcal J_1$ is a subset of $\{0,1,2,3,4\}$, and $\mathcal J_2$ is either empty if $D = 0$ or a subset of $\{1,\ldots,\min(D,3)\}$ if $D \geq 1$. We caution that in practice, assessing the validity of the monotonicity assumption for $D>2$ can be difficult unless in the presence of huge data size. Moreover, here we have assumed the moment exists if it is used, whereas in practice one may want to use the ratio-of-maximum-and-sum method \cite{embrechts2013modelling} to assess their finiteness.



\begin{figure}[h]
 \center
  \includegraphics{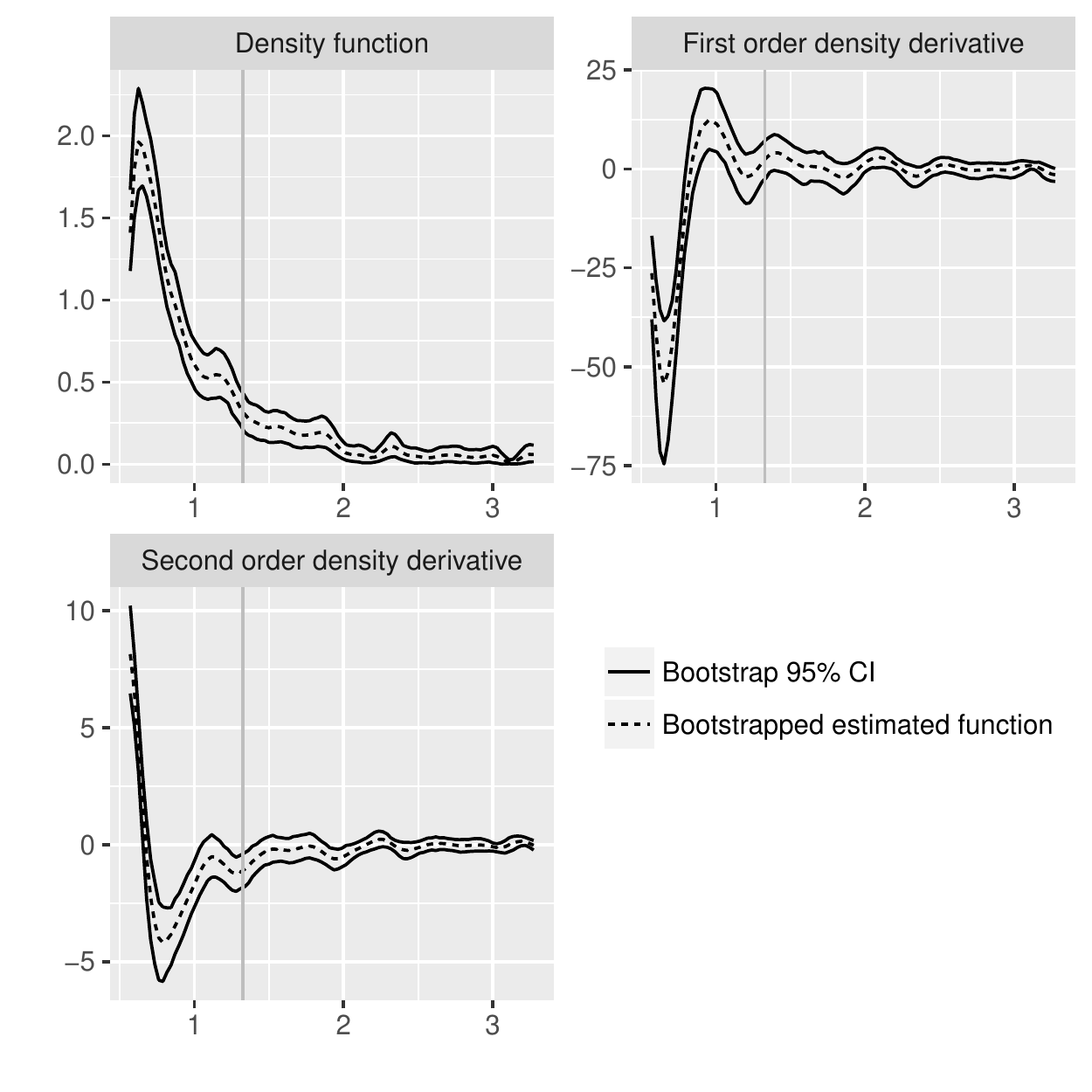} \caption{\label{fig : Params} \textit{Plots of the density and its derivative estimates. The dashed black lines represent the point estimates of the density function $f_X$ and its first two derivatives. The black lines show their $95\%$ confidence intervals. The gray lines locate our chosen threshold $a$.} }
 \end{figure}

To obtain a good choice of $a$ and the reliability of the constraints, we plot the trends of the density and density derivatives in Figure \ref{fig : Params}. The confidence intervals for various $a$'s are constructed using the bootstrap with $1000$ replicates on the built-in kernel estimates in R \cite{wand1994kernel}. We pick $a=1.35$, which is roughly the $80$-percentile of the data. Beyond this value (shown by the gray vertical line), the density function can be seen to be non-decreasing and convex. The signs of the estimates of the density and its derivative, however, are not as clear. A risk-averse user in this case would use $D=2$, and set $\mathcal J_2=\{1\}$.


We calibrate the normal confidence intervals for $E_F[(X-a)_+^j]$. To account for simultaneous estimation, we apply a Bonferroni correction in constructing these intervals together with those for $F^{(j)}(a)$.



We apply Corollary \ref{corly:moment LP} and Algorithm \ref{algo:GKLP} to solve Program  \eqref{opt : original inf const numerics} for all possible combinations of $p \in  \{90,99,99.9,99.99\}$,  $\mathcal J_1 \subseteq \{0,1,2,3,4\}$, $D \in \{0,\ldots,5\}$, $\mathcal J_2 \subseteq \{1,2,\min(3,D)\}$ if $D \geq 1$ and $\mathcal J_2 = \emptyset$ if $D = 0$. For $D>1$, we use numerical differentiation on the density estimates and apply the same bootstrap calibration procedure described before. 

\begin{figure}[h]
 \center
  \includegraphics[scale=0.6]{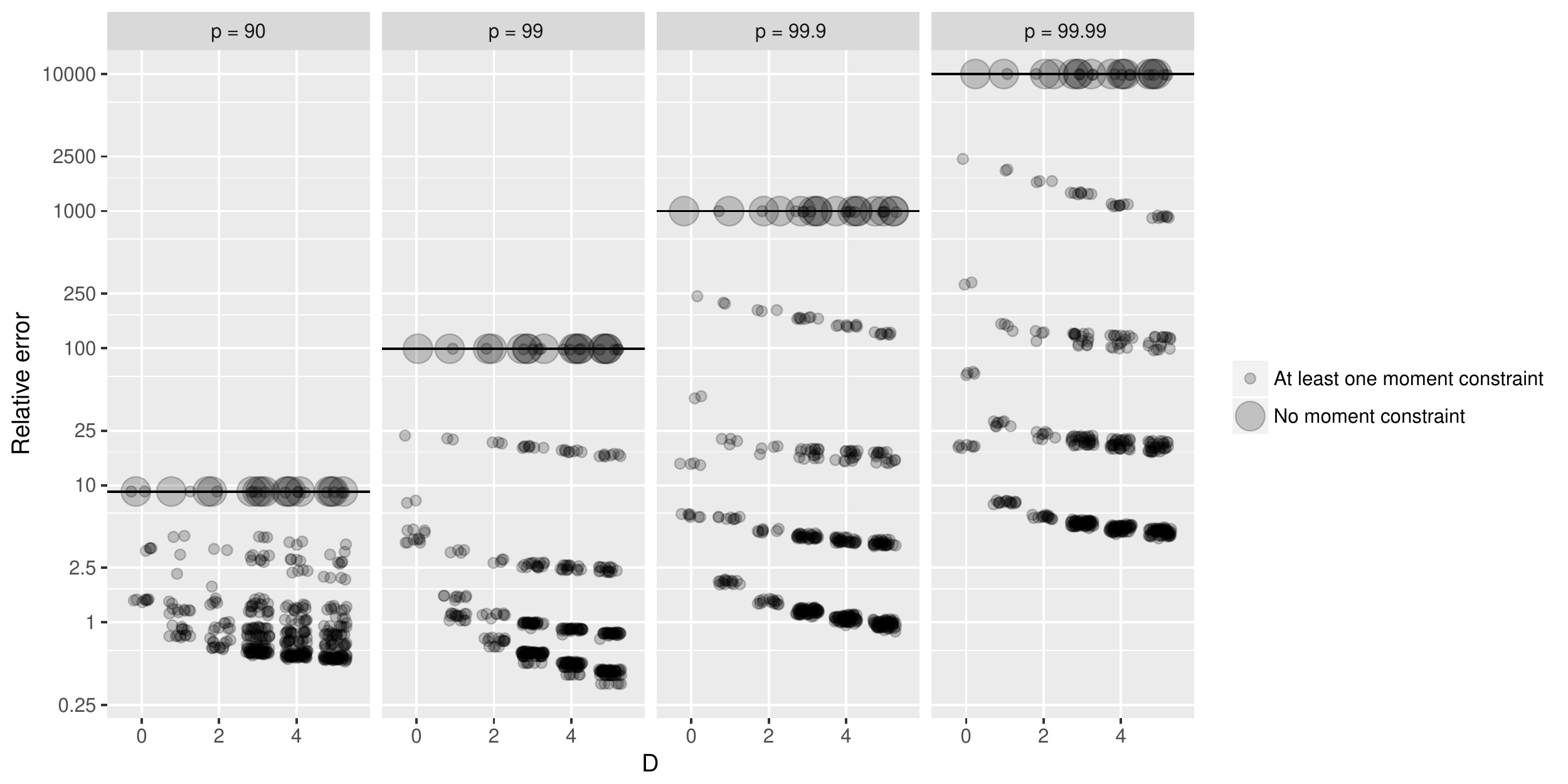} \caption{\label{fig : Bounds Horror} \textit{Relative error (defined as the relative increase of the optimization output value over the truth) for different $p$ and choices of $D$ and other constraints in the optimization. The solid lines represent the maximal relative error for a given value $p$.}}
 \end{figure}

The results are displayed in Figure \ref{fig : Bounds Horror}. Each point gives, for a given combination of the parameters $D$, $p$, and the sets $\mathcal J_1$ and $\mathcal J_2$, the relative error between the output of Program \eqref{opt : original inf const numerics} and the true value of $P(X \geq q_p)$. For a given $p$, the smallest relative error decreases with $D$, as the intuition suggests. In addition, the smallest relative error across all $D$ values increases with $p$. This can be attributed to the fact that the non-tail data are less informative as we infer on quantities associated with farther part of the tail. The large circles show the output with no moment constraints (in particular, without $\underline\gamma_{0,1}\leq \bar F(a)\leq\overline\gamma_{0,1}$ discussed in Section \ref{sec:overview}), which can be shown to give extremely conservative bounds.

Table \ref{Tab:best subset example} shows, for each given $D$ and each of the four values of $p$, the sets $\mathcal J_1^*$ and $\mathcal J_2^*$ giving the smallest relative error obtained across all possible combinations of the sets $\mathcal J_1$ and $\mathcal J_2$ tested (as shown in Figure \ref{fig : Bounds Horror}).  The relative error is defined as the relative increase of the optimization output value over the truth. This is from one set of data (without replication), so the value of relative error may vary. However, it can show some general pattern. In all four cases, the optimization outputs appear to capture the order of magnitude of the true underlying probability. The relative error in the case $D=0$ seems to be significantly larger than using at least $D=1$ (i.e., monotonicity of the tail density). The gain in relative error decreases as $D$ increases to 2 (i.e., convexity of the tail density). For $p=90$- or $99$-percentile, the relative error when using $D=2$ and one or two moment constraint is kept at a decimal. For $p=99.9$- or $99.99$-percentile, the relative error is larger, but encouragingly, it is still within a single digit. The table also shows that increasing $D$ further does not result in dramatic improvement. This suggests that adding a monotonicity constraint on higher order derivatives without  including a bound on the derivative itself is negligible. 


 \begin{table}[hp]
 \center
\begin{subtable}{\textwidth}
\centering
\begin{tabular}{|c|c|c|c|c|}
  \hline
$D$ & $\mathcal J_1^*$ & $\mathcal J_2^*$ & Optimal Objective Value & Relative Error \\ 
  \hline
0 & \{0\} & \{\} & 2.4e-01 & 1.401 \\ 
  1 & \{0, 1\} & \{\} & 1.78e-01 & 0.777 \\ 
  2 & \{0, 1\} & \{\} & 1.64e-01 & 0.635 \\ 
  3 & \{0, 1\} & \{\} & 1.57e-01 & 0.573 \\ 
  4 & \{0, 1\} & \{\} & 1.54e-01 & 0.539 \\ 
  5 & \{0, 1\} & \{\} & 1.52e-01 & 0.516 \\ 
   \hline
\end{tabular}\subcaption{$p = 90 $}
\end{subtable}

\begin{subtable}{\textwidth}
\centering
\begin{tabular}{|c|c|c|c|c|}
  \hline
$D$ & $\mathcal J_1^*$ & $\mathcal J_2^*$ & Optimal Objective Value & Relative Error \\ 
  \hline
0 & \{1, 3\} & \{\} & 4.82e-02 & 3.817 \\ 
  1 & \{0, 3\} & \{\} & 2.03e-02 & 1.032 \\ 
  2 & \{0, 3\} & \{\} & 1.66e-02 & 0.665 \\ 
  3 & \{0, 3\} & \{\} & 1.5e-02 & 0.505 \\ 
  4 & \{0, 3\} & \{\} & 1.41e-02 & 0.414 \\ 
  5 & \{0, 3\} & \{\} & 1.36e-02 & 0.356 \\ 
   \hline
\end{tabular}\subcaption{$p = 99 $}
\end{subtable}

\begin{subtable}{\textwidth}
\centering
\begin{tabular}{|c|c|c|c|c|}
  \hline
$D$ & $\mathcal J_1^*$ & $\mathcal J_2^*$ & Optimal Objective Value & Relative Error \\ 
  \hline
0 & \{1, 3\} & \{\} & 6.86e-03 & 5.856 \\ 
  1 & \{1, 3\} & \{\} & 2.89e-03 & 1.887 \\ 
  2 & \{1, 3\} & \{\} & 2.35e-03 & 1.349 \\ 
  3 & \{1, 3\} & \{1\} & 2.1e-03 & 1.098 \\ 
  4 & \{1, 3\} & \{1\} & 1.95e-03 & 0.952 \\ 
  5 & \{1, 3\} & \{1\} & 1.86e-03 & 0.856 \\ 
   \hline
\end{tabular}\subcaption{$p = 99.9 $}
\end{subtable}

\begin{subtable}{\textwidth}
\centering
\begin{tabular}{|c|c|c|c|c|}
  \hline
$D$ & $\mathcal J_1^*$ & $\mathcal J_2^*$ & Optimal Objective Value & Relative Error \\ 
  \hline
0 & \{2, 3\} & \{\} & 1.97e-03 & 18.688 \\ 
  1 & \{2, 3\} & \{\} & 8.26e-04 & 7.259 \\ 
  2 & \{1, 3\} & \{1\} & 6.66e-04 & 5.658 \\ 
  3 & \{1, 3\} & \{1\} & 5.86e-04 & 4.862 \\ 
  4 & \{1, 3\} & \{1\} & 5.4e-04 & 4.397 \\ 
  5 & \{1, 3\} & \{1\} & 5.09e-04 & 4.092 \\ 
   \hline
\end{tabular}\subcaption{$p = 99.99 $}
\end{subtable}

\caption{\label{Tab:best subset example}\textit{For each one of the five values of $D$ and for all four values of $p$, the sets $\mathcal J_1^*$ and $\mathcal J_2^*$ are the sets giving the smallest relative error obtained across all possible combinations of the sets $\mathcal J_1$ and $\mathcal J_2$ tested.}}
 \end{table}


We conclude this section by pointing out some subtle numerical issues when implementing the algorithm, related to the choice of $C$ in the compact-support moment problem formulation \eqref{opt: lambda infty finite support}. Note that $C$ is not known in the specification and needs to be chosen through trial and error. In our implementation, for all cases choosing $C$ in the tens work. However, one may encounter other examples in which $C$ needs to be chosen much higher. For instance, if we consider using $e^X$ instead of $X$ in our current data set, we found that the proper $C$ is in the range of thousands. This could cause numerical instability in R, but could potentially be well-implemented with more powerful optimization software.

\section{Conclusion}\label{sec:conclusion}
We have investigated an optimization-based approach to bound expectation-type extremal performance measures. The approach utilizes constraints to encode information about the monotonicity-type behaviors of the tail and moments and aims to compute the worst-case value among all tail distributions subject to these constraints. We have developed two results, one on the transformation from monotonicity constraints to moment constraints by using elementary integration by parts and change of measures, and show that in general there can be multiple equivalent transformed formulations. We have also developed a method to transform an infinite-support moment problem into a compact-support moment problem, which avoids non-convergence issues when running techniques like generalized linear programming due to escaping masses arising from the infinite support. A numerical example demonstrates the application of our approach and theoretical results. 

\section*{Acknowledgements}
We gratefully acknowledge support from the National Science Foundation under grants CMMI-1542020, CMMI-1523453 and CAREER CMMI-1653339.

\bibliographystyle{informs2014}
\bibliography{bibliography}

\ECSwitch


\ECHead{Appendix}

\section{Technical Proofs for Section \ref{sec : SILP}} \label{sec : technical proofs}

\proof{Proof of Lemma \ref{lemma : property set PL_new}:} Let $D \in \mathbb N^*$ and $F \in \mathcal P^D[a,\infty)$. The proof  focuses on the case when $a = 0$; this is without loss of generality since $F\in\mathcal P^{D}[a,\infty)$ if and only if $F(\cdot+a)\in\mathcal P^{D}(\mathbb R^+)$. \\

\noindent \textit{Item \ref{lemma: item derivative to 0}:} 
Since $F_+^{(D)}$  is monotone, it has a limit, say $l$, at $\infty$. If $l > 0$, the function $F^{(D-1)}$ is ultimately increasing, and there exists $K \in \mathbb R^+$ such that
\begin{equation}
F^{(D-1)}(x) - F^{(D-1)}(K) =\int_K^x F_+^{(D)}(u)du \to\infty \ \text{as} \ x \to \infty
\end{equation}
Hence, $\lim_{x\to\infty} F^{(D-1)}(x) = \infty$. One can repeat this argument to show  that $F^{(j)}$ is ultimately increasing and $\lim_{x\to\infty} F^{(j)}(x) = \infty$ for all $j \in \{0, D-1\}$. This  is  a contradiction as $F$ is bounded. In the same way, we can also prove that the limit $l$ cannot be negative, for otherwise $F$ would be ultimately decreasing. We therefore conclude that $l= 0$.\\

\noindent\textit{Item \ref{lemma: item non-decreasing set}:} By definition, the function $F_+^{(D)}$ is either non-increasing or non-decreasing, and converges to $0$ as shown in Item \ref{lemma: item derivative to 0}. As such, it never changes sign and $F^{(D-1)}$ is monotone as well. Proceeding by induction, we obtain $$\mathcal P^D(\mathbb R^+) \subset  \mathcal P^{D-1}(\mathbb R^+) \subset \ldots \subset \mathcal P^0(\mathbb R^+)  = \mathcal P(\mathbb R^+)$$

\noindent\textit{Item \ref{lemma: item alterning monotonicity}:} Building upon the proof of Item \ref{lemma: item non-decreasing set}, we see that if  $F_+^{(D)}$ is non-decreasing, it is non-positive and so $F^{(D-1)}$ is non-increasing and non-negative. Similarly, if  $F_+^{(D)}$ is non-increasing (non-negative) then $F^{(D-1)}$ is non-decreasing (non-positive).  Repeating this logic, the derivatives of odd order $j$ must have the same sign as  $F_+^{(1)}$ which we know to be non-negative. \hfill \Halmos\\
\endproof
\begin{remark}\label{rmk: lemma holds for Q}
It is straightforward to see that the proofs and hence the statements in Lemma \ref{lemma : property set PL_new} hold in the more general case where $F$ belongs to $\mathcal Q^D[a,\infty)$ defined in Section \ref{sec : SILP}.
\end{remark}


%

The proof of Theorem \ref{theorem : reformulation SILP} requires the following three propositions.

\begin{proposition}\label{prop : monotone derivatives}
Let $D \in \mathbb N$, $g$ be a locally integrable function on $\mathbb R^+$ that is either bounded below or above, and $F \in \mathcal Q^D(\mathbb R^+)$. Then for any $j \in \{0, \ldots, D\}$, we have

\begin{equation}\label{eq : F to P}
\int_0^\infty g dF = \int_0^\infty g^{(-j)} d P^{(j)}
\end{equation}
where $P^{(j)}(x) = (-1)^j F_+^{(j)}(x)$ for all $x\in \mathbb R^+$.
\end{proposition}

\proof{Proof of Proposition \ref{prop : monotone derivatives}:}
The proposition trivially holds for $D = 0$ so we focus on the case $D \geq 1$. First,  we  establish the validity of the statement  when $g$ is non-negative and $D = 1$. We then  generalize the result to unsigned functions that are either bounded above or below. The conclusion will hold for any $D>1$ by recursing the argument.\\

\noindent \textit{Step 1:} Because the function $g$ is locally integrable, its first order antiderivative  $g^{(-1)}$ exists and is continuous. In addition, the right derivative $F_+^{(1)}$ is of bounded variation.  The integral $\int_0^x g^{(-1)} dF_+^{(1)}$ therefore exists for all  $x \geq 0$ and, with an integration by part, we obtain that for all $x\in \mathbb R^+$
\begin{align}
\int_{0}^x g^{(-1)} dF_+^{(1)} + \int_0^x  F_+^{(1)} dg^{(-1)} &= g^{(-1)}(x) F_+^{(1)}(x)  = - g^{(-1)}(x) \int_x^\infty d F_+^{(1)}
\label{eq : g-d+1 to g-d}
\end{align}
where the last equality is a consequence of Lemma \ref{lemma : property set PL_new}.\ref{lemma: item derivative to 0}. Because $g$ is non-negative, the function  $g^{(-1)}$ is  non-decreasing and non-negative. Hence,
\begin{equation}\label{eq:pos g inc  pos G}
 0 \leq - g^{(-1)}(x) \int_x^\infty  d F_+^{(1)}  \leq  - \int_x^\infty g^{(-1)}d F_+^{(1)}, \ \ \ \text{for all } x\in \mathbb R^+
\end{equation}

In addition,
\begin{align}
\int_0^x  F_+^{(1)} dg^{(-1)} &= \int_0^x  F_+^{(1)}(u) g(u) du= \int_0^x  g d F \label{eq F+1dg1 = g0 dF0}
\end{align}
Combining \eqref{eq : g-d+1 to g-d} and \eqref{eq:pos g inc  pos G}, we have
\begin{equation}
0 \leq\int_{0}^x g^{(-1)} dF_+^{(1)} + \int_0^x  F_+^{(1)} dg^{(-1)}= - g^{(-1)}(x) \int_x^\infty  d F_+^{(1)}  \leq  - \int_x^\infty g^{(-1)}d F_+^{(1)}, \ \ \ \text{for all } x\in \mathbb R^+\label{interim new}
\end{equation}
Subtracting $\int_0^xg^{(-1)}dF_+^{(1)}$ in \eqref{interim new} gives
$$ -\int_{0}^x g^{(-1)} dF_+^{(1)} \leq \int_0^x  F_+^{(1)} dg^{(-1)}\leq   -\int_{0}^x g^{(-1)} dF_+^{(1)} - \int_x^\infty g^{(-1)}d F_+^{(1)}, \ \ \ \text{for all } x\in \mathbb R^+$$
or
\begin{equation}
-\int_{0}^x g^{(-1)} dF_+^{(1)} \leq \int_0^x  F_+^{(1)} dg^{(-1)}\leq   -\int_{0}^\infty g^{(-1)} dF_+^{(1)}, \ \ \ \text{for all } x\in \mathbb R^+\label{interim new1}
\end{equation}
Substituting \eqref{eq F+1dg1 = g0 dF0} into \eqref{interim new1}, we have
\begin{equation}\label{eq : g-d+1 to g-d ineq F_+}
-\int_{0}^x g^{(-1)} dF_+^{(1)} \leq \int_0^x  g d F\leq   -\int_{0}^\infty g^{(-1)} dF_+^{(1)}, \ \ \ \text{for all } x\in \mathbb R^+
\end{equation}

Taking the limit on both sides of  \eqref{eq : g-d+1 to g-d ineq F_+}, we obtain \eqref{eq : F to P}.\\

\noindent \textit{Step 2:} We now consider the case when $g$ is an unsigned function bounded below by a constant $m \in \mathbb R$.  Inequality \eqref{eq : F to P} then applies for both the function $g - m$ and the constant function $1$. As a result,  equalities $\int_0^\infty (g - m) dF= \int_0^\infty (g^{(-1)} - mx) dP^{(1)}$ and  $\int_0^\infty d F = \int_0^\infty x dP^{(1)}$  hold. In fact,  the last equality is bounded since $F \in \mathcal Q(\mathbb R^+)$, and

\begin{align}
\int_0^\infty g dF &= \int_0^\infty(g - m)dF + m \int_0^\infty dF \nonumber\\
& = \int_0^\infty (g^{(-1)} - m x)dP^{(1)} + m \int_0^\infty x dP^{(1)}\nonumber \\
&=  \int_0^\infty g^{(-1)} dP^{(1)}
\end{align}

When $g$ is an unsigned function bounded above by  a constant  $M \in \mathbb R$.  Inequality \eqref{eq : F to P} then applies for the function $(M - g)$ and $- \int_0^\infty g dF  = \int_0^\infty (M - g) - M  dF = - \int_0^\infty g^{(-1)}dP^{(1)}$. This concludes our proof.\hfill \Halmos\\
\endproof

\begin{proposition}\label{prop : monotone derivatives 2}
Let $D \in \mathbb N\setminus\{0\}$, $F \in \mathcal Q^D(\mathbb R^+)$,  and $P(x) = (-1)^D F_+^{(D)}(x)$ for all $x \in \mathbb R^+$.  Then,

\begin{enumerate}
\item  $(-1)^{(j+1)} F_+^{(j)}(0) =\int \frac{u^{D-j}}{(D-j)!}d P(u)$  for all  $j \in \left\lbrace 1, \ldots, D\right\rbrace$
\item $\lim_{x\to\infty}F(x)= \int \frac{u^{D}}{D! } d P(u)$
\end{enumerate}
\end{proposition}

\proof{} By definition of  the function $P^{(j)}$ defined in Proposition \ref{prop : monotone derivatives}, we have

$$\lim_{x\to \infty }(-1)^j\left[  F_+^{(j)}(x) - F_+^{(j)}(0)\right]  =\int_0^\infty dP^{(j)}(u)$$

 Applying Proposition \ref{prop : monotone derivatives} with $g(u)\equiv 1$ also gives

\begin{align}\label{eq : monotone derivatives 2 main equality}
\lim_{x\to \infty }(-1)^j\left[  F_+^{(j)}(x) - F_+^{(j)}(0)\right] &=  \int_0^\infty g^{(-(D-j))}dP^{((D-j)+j)} = \int_0^\infty \frac{u^{D-j}}{(D-j)!}dP(u)
\end{align}

The first item then follows from  $\lim_{x\to\infty}F_+^{(j)}(x)= 0$ for all $j \in \{1,\ldots, D \}$ by Lemma \ref{lemma : property set PL_new}.\ref{lemma: item derivative to 0}. The second item is a consequence of the  continuity of $F$ when $D \geq 1$  and the definition of $\mathcal Q^D(\mathbb R^+)$ that $F(0)=0$.\hfill \Halmos
\endproof

\begin{proposition}\label{prop: F to P}
 Let $D \in \mathbb N\setminus\{0\}$. A function $F$ is an element of $\mathcal Q^D(\mathbb R^+)$ if and only if there exists a function $Q$ such that $Q(x) \in \mathcal Q(\mathbb R^+)$, $Q(0) = 0$, $\int x^D dQ(x) < \infty$, and

\begin{equation}\label{eq : Pl reform }
 D! F(x) = \int  \left(u^D - (u-x)^{D}I(u > x) \right) dQ(u) \ \ \mbox{for all } x \in \mathbb R
\end{equation}
In fact, $Q(x) = (-1)^{D} \left[F^{(D)}(x) - F^{(D)}(0)\right]$ for all $x \in \mathbb R^+$.
\end{proposition}

\proof{}
We first show that any function $F \in \mathcal Q^D(\mathbb R^+)$ can be expressed in the form of \eqref{eq : Pl reform }. Let $h(u,x)$ be the function defined as $D! h(u,x) = u^D - (u-x )^{D}I(u > x)$. Then for all $(u,x) \in \mathbb R^+ \times \mathbb R^+$ and $j \in \{0,\ldots, D\}$,
\begin{align}
\frac{\partial^j h}{\partial u^j}(u,x) &=\frac{1}{(D-j)! } \left[u^{D - j} - (u-x)^{D-j}I(u > x)\right]
\end{align}

In particular, $\frac{\partial^j h}{\partial u^j} (0,x) = 0$ for all $j \in \{ 0,\ldots, D-1\}$ and $\frac{\partial^D h}{\partial u^D} (u,x) = I(u \leq x)$. The function $h(u,x)$ is therefore the $D^{th}$ order anti-derivative with respect to $u$ vanishing at $0$ of the function $I(u \leq x)$.  Since $F(x) = \int I(u\leq x) dF(u)$, an application of Proposition \ref{prop : monotone derivatives} shows that the distribution function $F$ can be written as
\begin{align*}
F(x) &= \int h(u,x)dP(u)
\end{align*}
where $P(x) = (-1)^{D}F_+^{(D)}(x) $ for all $x\in \mathbb  R^+$. Hence, $Q(x)  = P(x) - P(0)$ and $Q(0) = 0$ trivially holds. From Lemma \ref{lemma : property set PL_new}.\ref{lemma: item alterning monotonicity} and Remark \ref{rmk: lemma holds for Q}, we also have $Q(x)  \in \mathcal Q(\mathbb R^+)$.  In addition, $F(x) = \int h(u,x)dP(u) =\int h(u,x)dQ(u)$. Lastly, the integral $\int u^D/D! dP(u)$ is bounded since it is the limit of the distribution function $F$ by Proposition \ref{prop : monotone derivatives 2}, and hence so is $\int u^D dQ(u)$.\\

We now focus on the other direction of the statement, $i.e.$ we consider the case when $Q$ is a distribution function on $\mathbb R^+$ satisfying $\int u^D dQ(u) < \infty$, $Q(0) = 0$, and $F$ is as defined in \eqref{eq : Pl reform }. The function $F$ is then absolutely continuous and non-decreasing. Moreover, we have $(D-1)! F_+^{(1)}(x) = \int (u-x)^{D-1} I(u > x) dQ(u)$ for all $x \in \mathbb R^+$ by an interchange of derivative and the integral, justified since $u^{D} -(u-x)^{D} I(u > x)$ is $Q$-integrable and  $(u-x)^{D-1} I(u > x)$ is bounded by a $Q$-integrable function, which in turn is guaranteed since $\int u^DdQ<\infty$ (e.g., Theorem 6.28 in \cite{klenke2013probability}).



Iteratively, we obtain $(-1)^{D+1}F_+^{(D)}(x) = \int I(u > x) dQ(u)$ for all $x\in \mathbb R^+$.  Hence, $(-1)^{D}\left[F_+^{(D)}(x) - F_+^{(D)}(0)\right] = Q(x) - Q(0) = Q(x)$. As $Q$ belongs to $\mathcal Q(\mathbb R^+)$, the function $F_+^{(D)}$ is monotone. It remains to show that $F$ is bounded to conclude that $F\in\mathcal Q^D(\mathbb R^+)$.  Because $F$ is non-decreasing, it is enough to prove that its limit is bounded. By definition, the $D^{th}$ moment of $Q$ is bounded. Therefore, using the monotone convergence theorem, we have
\begin{align*}
\lim_{x \to \infty}D! F(x) = \int  \lim_{x \to \infty} \left[u^D - (u-x)^{D}I(u > x) \right] dQ(u) = \int u^D dQ(u) <\infty
\end{align*}
\hfill \Halmos
\endproof

\proof{Proof of Theorem \ref{theorem : reformulation SILP}:} The distribution function $F$ is an element of $\mathcal Q^D[a,\infty)$ if and only if $F(x+a)$ is an element of $\mathcal Q^D(\mathbb R^+)$. When $D = 0$,  the set $\mathcal J_2$ is empty by definition and program \eqref{opt : original inf const} can be reformulated as  \eqref{opt : SILP} by applying the change of variables $\mathcal J = \mathcal J_1$, $P =  F \circ u_a$,  $H  =  h \circ u_a$, and $G_j = g_{j_1} \circ u_a$ for all $j \in \mathcal J_1$, so that the conclusion holds. In the remainder of this proof, we focus on the case $D \geq 1$.

With Assumption \ref{assump:h},  Proposition \ref{prop : monotone derivatives} allows us to reformulate the objective value and the first set of inequality constraints as follows:
\begin{align}
&\int h(x) dF(x) = \int h(x+a) dF(x+a) = \int (h \circ u_a)^{(-D)} d P\label{eq : transform objective}\\
&\int g_{j,1}(x) dF(x) = \int g_{j,1}(x+a)dF(x+a) = \int  (g_j \circ u_a)^{(-D)} d P, \ \ \text{for all $j \in \mathcal J_1$}
\end{align}
where $P(x) = (-1)^D F_+^{(D)}(x+a) $  for all $x \in \mathbb R^+$. Moreover, Proposition \ref{prop : monotone derivatives 2} implies that
\begin{align}
(-1)^{j+1}F_+^{(j)}(a) = \int \frac{x^{D-j}}{(D-j)!}dP, \ \ \text{for all $j \in \mathcal J_2$}  \label{eq : transform constraints}
\end{align}


So \eqref{opt : original inf const} is the same as
\begin{equation}
\begin{array}{lll}
\underset{F}{\sup}&\int  (h \circ u_a)^{(-D)} dQ\\
\text{subject to\ \ }
& \int  (g_j \circ u_a)^{(-D)} dQ \leq \gamma_{j,1} &\text{\ \ for all $j \in \mathcal J_1$}\\
&\underline \gamma_{j,2}\leq \int x^{D-j}/(D-j)! d Q \leq \overline \gamma_{j,2} &\text{\ \ for all $j \in \mathcal J_2$}\\
& Q(x) = (-1)^D \left[F_+^{(D)}(x+a) - F_+^{(D)}(a)\right]&  \text{\ \ for all $x \in \mathbb R$}\\
& F(x+a) \in \mathcal Q^D(\mathbb R^+)
\end{array}\label{opt : SILP derivative}
\end{equation}

By Proposition \ref{prop: F to P}, $F(\cdot+a)\in\mathcal Q^{D}(\mathbb R^+)$ if and only if there exists $Q(\cdot) \in \mathcal Q(\mathbb R^+)$ such that $Q(0) = 0$, $\int x^D dQ(x)$ is bounded, and
\begin{equation}\label{eq:link F to P}
D! F(x+a) = \int u^D - (u - x)^DI(u > x) dQ(u)
\end{equation}
for all $x \in \mathbb R$. Moreover, $Q(\cdot)$ must satisfy $Q(x) = (-1)^D \left[F_+^{(D)}(x+a) - F_+^{(D)}(a)\right]$ for all $x \in \mathbb R$.  Therefore, we obtain that  \eqref{opt : SILP derivative} is the same as
\begin{equation}
\begin{array}{lll}
\underset{P}{\sup}&\int (h \circ u_a)^{(-D)}  dQ\\
\text{subject to\ \ }
& \int (g_j \circ u_a)^{(-D)}  dQ \leq \gamma_{j,1} &\text{\ \ for all $j \in \mathcal J_1$}\\
&\underline \gamma_{j,2}\leq \int x^{D-j}/(D-j)!  d Q(x) \leq \overline \gamma_{j,2} &\text{\ \ for all $j \in \mathcal J_2$}\\
& \int x^D dQ<\infty\\
& Q(0) = 0\\
& D! F(x+a) = \int u^D - (u - x)^DI(u > x) dQ(u)\\
& Q \in \mathcal Q(\mathbb R^+)
\end{array}\label{opt : SILP almost}
\end{equation}
The last inequality constraint $\int x^DdQ<\infty$ is redundant and can be dropped, since $\bar F(a) = \int x^D/D! dQ(x) \leq \overline \gamma_{0,1}$ must be one of the constraint of Program \eqref{opt : SILP almost} by construction. Moreover, since $ (h \circ u_a)^{(-D)}(0) = (g_j \circ u_a)^{(-D)}(0) = 0$, the constraint  $Q(0) = 0$ impacts neither the feasible set nor the objective value and can also be dropped. Hence,  \eqref{opt : SILP almost} is the same as
\begin{equation}
 \begin{array}{lll}
Z^* = \underset{P}{\sup}&\int (h \circ u_a)^{(-D)}  dQ \\
\text{subject to\ \ }& \int (g_j \circ u_a)^{(-D)}   dQ\leq \gamma_{j,1} &\text{\ \ for all $j \in \mathcal J_1$}\\
& \underline \gamma_{j,2}\leq \int x^{D-j}/(D-j)!  dQ\leq \overline \gamma_{j,2} &\text{\ \ for all $j \in \mathcal J_2$}\\
& D! F(x+a) = \int u^D - (u - x)^DI(u > x) dQ(u)\\
& Q \in \mathcal Q(\mathbb R^+)
\end{array}\label{opt : SILP almost 2}
\end{equation}

Since $\int x^DdQ<\infty$, we can define a distribution function $\tilde Q \in \mathcal Q(\mathbb R^+)$ absolutely continuous with respect to $Q$ via $d\tilde Q=x^{D-J}dQ$, i.e., the Radon-Nikodym derivative given by $\frac{d\tilde Q}{dQ}=x^{D-J}$, where $J$ can be taken as any integer in $\{0,\ldots,D\}$. Converting the decision variable from $Q$ to $\tilde Q$ in \eqref{opt : SILP almost 2} gives
\begin{equation}
 \begin{array}{lll}
Z^* = \underset{P}{\sup}&\int H d\tilde Q  \\
\text{subject to\ \ }& \int G_{j,1}  d\tilde Q \leq \gamma_{j,1} &\text{\ \ for all $j \in \mathcal J_1$}\\
& \underline \gamma_{j,2}\leq \int G_{j,2} d\tilde Q  \leq \overline \gamma_{j,2} &\text{\ \ for all $j \in \mathcal J_2$}\\
& D! F(x+a) = \int \left[u^D - (u - x)^DI(u > x)\right]u^{J-D} d\tilde Q(u)\\
& \tilde Q \in \mathcal Q(\mathbb R^+)
\end{array}\label{opt : SILP almost 3}
\end{equation}

The constraint defining the function $F$ does not affect \eqref{opt : SILP almost 3}. Therefore,  \eqref{opt : original inf const} and \eqref{opt : SILP not condensed}  have the same optimal value, and if $\tilde Q^*$ is an optimal solution of \eqref{opt : SILP not condensed}, then the function $F^*$ defined as
\begin{equation}\label{eq:link F to Q}
D! F^*(x+a) = \int u^J\left(1 - (1 - x/u)^D I(u > x)\right) d\tilde Q^*(u)
\end{equation}
is an optimal solution of \eqref{opt : original inf const}.\hfill \Halmos
\endproof
\section{Technical Proofs for Section \ref{sec : formulation LP}}\label{sec : appendix tech proof part 2}

\proof{Proof of Theorem \ref{theorem:include lambda 0 infty}:}
Let $x \in  \mathbb R^+ \cup \{\infty\}$.  By Assumption \ref{slater}, Theorem \ref{theorem:strong duality} and Remark \ref{rmk: strong duality} below, strong duality holds and we have
\begin{equation}
\begin{array}{lllll}
Z^* = \underset{y}{\inf} &\sum_{j \in \mathcal J}  y_j \gamma_j  \\
\text{subject to}&\sum_{j \in \mathcal J}  y_j G_j(u) \geq H(u)&\text{\ \ for all\ }u\in \mathbb R^+\\
&y_j \geq 0& \text{\ \ for all\ } j \in \mathcal J
\end{array}\label{opt:SILP dual b}
\end{equation}
which is the dual formulation of \eqref{opt : SILP}. Suppose that $\mathcal J(x)$ is non-empty. Then, we consider the sequence $u_n$ in the definition of $\mathcal J(x)$. For all $i \in \mathcal J(x)$,  $u_n \in \text{supp}(G_i)$ and \eqref{opt:SILP dual b} satisfies the  implicit constraint
\begin{equation}
\limsup_{u_n \to x}\left(\sum_{j\in \mathcal J} y_j\frac{G_j(u_n)}{|G_i(u_n)|}\right) \ \geq \ \limsup_{u\to x}\frac{H(u_n)}{|G_i(u_n)|}\label{ineq: SILP dual implied const}
\end{equation}
By the definition of $\mathcal J(x)$, \eqref{ineq: SILP dual implied const} gives
\begin{equation}
 0 \leq  \limsup_{u_n\to x}\frac{H(u_n)}{|G_i(u_n)|} \leq  \sum_{j \in \mathcal J \setminus \left\lbrace i\right\rbrace}  y_j \limsup_{u_n\to x}\frac{G_j(u_n)}{|G_i(u_n)|} - y_i\leq  -y_i \leq 0 \label{ineq:necessary cond dual}
\end{equation}

As a consequence,   the dual multipliers $y_i$'s must be 0 for all $i \in \mathcal J(x)$. \eqref{opt:SILP dual b} is then equivalent to

\begin{equation}
\begin{array}{lllll}
Z^* = \underset{y}{\inf} &\sum_{j \in \mathcal J\setminus \mathcal J(x)}  y_j \gamma_j  \\
\text{subject to}&\sum_{j \in \mathcal J\setminus \mathcal J(x)} y_j G_j(u) &\geq H(u)&\text{\ \ for all\ }u\in \mathbb R^+\\
&y_j \geq 0 & \text{\ \ for all\ } j \in \mathcal J\setminus \mathcal J(x)
\end{array}\label{opt : SILP dual drop J}
\end{equation}

Suppose $\mathcal J =  \mathcal J(x)$. When $\sup\{H(u)|u \in \mathbb R^+\} > 0$, \eqref{opt : SILP dual drop J} becomes infeasible because of the constraint $0\geq H(u)\ \text{for all\ }u\in\mathbb R^+$. When $\sup\{H(u)|u \in \mathbb R^+\} \leq 0$ , then $Z^*=0$. The case $\mathcal J(x)\subsetneq\mathcal J$ follows from strong duality again.\hfill \Halmos\\


\endproof

\proof{Proof of Lemma \ref{rmk:empty J}:}
Since each constraint in program \eqref{formulation empty J} involves a  lower bound and an upper bound,  the set $\mathcal J(x)$ defined in \eqref{redundant set} becomes
\small
\begin{eqnarray}
\mathcal J(x) = \Bigg \lbrace i \in  \tilde{\mathcal J} \Bigg|\exists u_n\in\text{supp}(G_i)\text{\ s.t.\ } \limsup_{u_n\to x} \frac{H(u_n)}{|\tilde G_{i}(u_n)|}    \geq 0 \mbox{ and }\limsup_{u_n\to x} \frac{-\tilde G_{j}(u_n)}{|\tilde G_{i}(u_n)|} \leq 0 \ \mbox{ and }  \limsup_{u_n\to x} \frac{\tilde G_{j}(u_n)}{|\tilde G_{i}(u_n)|} \leq 0 \ \forall j \in \tilde{\mathcal J}  \Bigg\rbrace \nonumber\\
\end{eqnarray}
\normalsize

For a given function $\tilde G_i$, the last two inequalities cannot hold at the same time for $j = i$ since  $\limsup_{u_n\to x} \frac{\tilde G_{i}(u_n)}{|\tilde G_{i}(u_n)|}$ is either equal to $1$ or $-1$. Hence $\mathcal J(x)  = \emptyset$.\hfill \Halmos\\
\endproof

\proof{Proof of Theorem \ref{theorem:existence bounded measure}:}
Applying Theorem \ref{theorem:include lambda 0 infty}, if $\mathcal J = \mathcal J(\infty)$, we fall into the trivial scenarios of the theorem yielding the first two items of Theorem \ref{theorem:existence bounded measure}. Otherwise, the constraints whose index fall in the set $\mathcal J(\infty)$ can be dropped and Theorem \ref{theorem:strong duality} together with Remark \ref{rmk: strong duality} below gives
\begin{equation}
\begin{array}{lllll}
Z^* = \underset{y}{\inf} &\sum_{j \in \mathcal J \setminus \mathcal J(\infty)}  y_j \gamma_j  \\
\text{subject to}&\sum_{j \in \mathcal J \setminus \mathcal J(\infty)} y_j G_j(u) &\geq H(u)&\text{\ \ for all\ }u\in \mathbb R^+\\
&y_j &\geq 0 & \text{\ \ for all\ } j \in \mathcal J \setminus \mathcal J(\infty)
\end{array}\label{opt : SILP dual drop Jx Jinfty}
\end{equation}

Under Assumption 
\ref{assump:regularityinfty},  \eqref{opt : SILP dual drop Jx Jinfty} satisfies the implicit constraints
\begin{align}
 \sum_{j \in \mathcal J \setminus \mathcal J(\infty)}  y_j    \limsup_{u\to \infty} \frac{G_j(u)}{|G_M(u)|} &\geq \limsup_{u\to \infty} \left( \sum_{j \in \mathcal J \setminus \mathcal J(\infty)}  y_j   \frac{G_j(u)}{|G_M(u)|}\right) \geq   \limsup_{u\to \infty} \frac{H(u)}{|G_{M}(u)|} = \lambda_M \label{ineq: implicit const J}
\end{align}
When  $\lambda_{M} = \infty$, \eqref{ineq: implicit const J} deems any solution with $y_j\in\mathbb R$ infeasible, and hence $Z^* = +\infty$. In the remainder of this proof, we only consider the case when $\lambda_M$ is finite.  \\ 

Since \eqref{opt : SILP} is a feasible program, there exists a sequence of feasible solutions $P^{(k)}$ such that $\int HdP^{(k)} \to Z^*$, and because $P^{(k)}\in \mathcal Q(\mathbb R^+)$, the integral $\int dP^{(k)}$ is bounded for all $k \in \mathbb N$. Hence, $Z^* = \lim_{k\to\infty} Z_{k}$ where
\begin{equation}
 \begin{array}{*{5}l}
Z_k =\underset{P}{\sup}&\int H dP\\
\text{subject to}&\int dP = \nu^{(k)}\\
&  \int G_j dP   \leq \gamma_j\text{\ \ for all $j \in \mathcal J \setminus \mathcal J(\infty)$}\\
& P \in \mathcal Q(\mathbb R^+)
\end{array}\label{opt:finite LP Znuk}
\end{equation}
and $\nu^{(k)} = \int dP^{(k)} $. Based on  Theorem \ref{theorem:discrete  support} below, it is sufficient  to investigate the sequences $P^{(k)}$ with at most $N$ point supports where $N$ is the number of linearly independent functions in the set $\left\lbrace\left( G_j \right)_{j \in \mathcal J\setminus \mathcal J(\infty)}, 1\right\rbrace$, $i.e.$ $P^{(k)} \in  \mathcal Q_{N}(\mathbb R^+)$. The sequence  $P^{(k)}$ can then be represented by $N$ couples of point masses and point supports $(p_{i}^{(k)}, x_{i}^{(k)})$ where we assume without loss of generality that $x_{1}^{(k)} \leq \ldots \leq x_{N}^{(k)}$. In particular, the sequence of $\sum_{i=1}^N p_{i}^{(k)}$ is bounded since, by Assumption \ref{assump:postivity}, there exists some $j \in \mathcal J \setminus \mathcal J(\infty)$ such that $\inf_{x\in \mathbb R^+} G_j(x) > 0$ so that
\begin{equation}\label{ineq: bounded mass}
\inf_{i \in \{1,\ldots N\}} G_j\left(x_i^{(k)}\right) \sum_{i = 1}^N  p_i^{(k)}  \leq  \sum_{i =1}^{N}  p_i^{(k)} G_j\left(x_i^{(k)}\right)  \leq \gamma_j
\end{equation}

Next, we define  the sequence $s^{(k')}=\sum_{i \in \mathcal I} p_i^{(k')} \left|G_M\left(x_i^{(k')}\right)\right|$ where $\mathcal I$ is the set containing the indexes of the support points which are unbounded for some subsequence, $i.e.$
\begin{equation}
\mathcal I = \left\lbrace i \in 1,\ldots,N| \lim_{k\to\infty} x_i^{(k)} = \infty \mbox{ for some subsequence indexed by\ } k\in \mathbb N \  \right\rbrace
\end{equation}
Moreover, we define $k'$ as the index of the sequence associated with the smallest element in the set $\mathcal I$ if the latter is not empty and $k' = k$ otherwise. As defined, the sequence $s^{(k')}$ is bounded. To see this, note that by definition of the set $\mathcal J \setminus \mathcal J(\infty)$, and under Assumptions \ref{assump:regularityinfty} and \ref{assump:well defined}, there exists some $M\in\mathcal J \setminus \mathcal J(\infty)$ such that either $\lambda_{j,M} > 0$ for some $j \in \mathcal J \setminus \mathcal J(\infty)$ or $ \lambda_{M} < 0$. Furthermore, the quantity $\lambda_{j,M}$ is finite by the same assumptions so that when $\lambda_{j,M} > 0$, we have for all $\varepsilon_1 \in (0, \lambda_{j,M})$ and $k' $  large enough,
\begin{align}
\gamma_j &\geq     \sum_{i = 1}^{N}  p_i^{(k')} G_j\left(x_i^{(k')}\right)  \nonumber\\
& = \sum_{i \notin \mathcal I}  p_i^{(k')} G_j\left(x_i^{(k')}\right)+\sum_{i \in \mathcal I}  p_i^{(k')}  \left|G_M\left(x_i^{(k')}\right)\right| \frac{G_j\left(x_i^{(k')}\right)}{\left|G_M\left(x_i^{(k')}\right)\right|}  \nonumber\\
&\geq \sum_{i \notin \mathcal I}  p_i^{(k')} G_j\left(x_i^{(k')}\right)+ \inf_{i \in \mathcal I}\left(\frac{G_j\left(x_i^{(k')}\right)}{\left|G_M\left(x_i^{(k')}\right)\right|}\right) s^{(k')}  \nonumber\\
&\geq \sum_{i \notin \mathcal I}  p_i^{(k')} G_j\left(x_i^{(k')}\right)    + (\lambda_{j,M}- \varepsilon_1) s^{(k')}
\label{ineq:const inf s bounded}
\end{align}
where the last inequality follows from the fact that
\begin{align}
\lim_{k'\to\infty} x_i^{(k')} = \infty, \ \ & \text{for all } i \in \mathcal I\\
 \limsup_{k'\to\infty} x_i^{(k')} < \infty \ \ & \text{for all } i \notin \mathcal I
\end{align}

Hence, $\sum_{i \notin \mathcal I}  p_i^{(k')}G_j\left(x_i^{(k')}\right)$
is finite and \eqref{ineq:const inf s bounded} implies the boundedness of the  sequence  $ s^{(k')}$ when $\lambda_{j,M} > 0$ for some $j \in \mathcal J\setminus \mathcal J(\infty)$. When $\lambda_M < 0$, we can show with a similar argument as in the derivation of \eqref{ineq:const inf s bounded} that  for all  $\varepsilon_2 \in (0,-\lambda_M)$ and  $k'$ large enough,
\begin{equation}
\sum_{i =1}^{N}  p_i^{(k')}H\left(x_i^{(k')}\right) \leq \sum_{i\notin \mathcal I}  p_i^{(k')}H\left(x_i^{(k')}\right) +
s^{(k')} \begin{cases}
\left(\lambda_M + \varepsilon_2\right)  & \mbox{if } -\infty < \lambda_M \\
-\varepsilon_2 & \mbox{if }\lambda_M = -\infty
\end{cases} \label{ineq : H sup finite support}
\end{equation}

The LHS in \eqref{ineq : H sup finite support} is bounded below since it converges to $Z^*$ and \eqref{opt : SILP} is consistent. In addition, the sum in the RHS of \eqref{ineq : H sup finite support} is finite by Assumption \ref{assump:semi-cont H}. The sequence $s^{(k')}$ is therefore  bounded when $-\infty < \lambda_M < 0$. Last but not least, $s^{(k')}$ must be $0$ for all $k'$ large enough when $\lambda_M = -\infty$; otherwise, we could choose $\varepsilon_2$ arbitrarily large and have the RHS tend to $-\infty$ as $\varepsilon_2$ grows.

Consequently, we have shown that $s^{(k')}$ is bounded whether $\lambda_{j,M}>0$ or $\lambda_M<0$, so there exists a subsubsequence  $k''$  such that  \begin{equation}\label{eq:opt sol solve lemma}
\begin{array}{cll}
  \left (p_i^{(k'')},x_i^{(k'')}\right)&\to (p_i^*, x_i^*) & \text{for all } i \notin \mathcal I \\
 s^{(k'')} &\to    s^* & \text{where }  s^* = 0  \text{ if } \lambda_M = -\infty \\
 \end{array}
 \end{equation}
where $p_i^*$, $x_i^*$, and $s^*$ are non-negative finite quantities, and for all $j \in \mathcal J \setminus \mathcal J(\infty)$,
\begin{eqnarray}
\gamma_j & \geq& \lim_{k''\to\infty} \sum_{i = 1}^{N}  p_i^{(k'')} G_j\left(x_i^{(k'')}\right) \nonumber\\
&\geq & \sum_{i \notin \mathcal I}   \liminf_{k''\to\infty} \left( p_i^{(k'')} G_j\left(x_i^{(k'')}\right)\right) + \lim_{k''\to\infty} \left(\sum_{i \in \mathcal I}  p_i^{(k'')} G_j\left(x_i^{(k'')}\right)\right)   \nonumber \\
&=& \sum_{i \notin \mathcal I} p_i^{*}G_j\left(x_i^{*}\right) + \lambda_{j,M} s^{*}
\label{ineq : bounded support const}
\end{eqnarray}
where the last inequality is a consequence of $G_j$ being lower semi-continuous by Assumption \ref{assump:semi-cont G}. We have therefore shown that $(P^*,  s^* )$, where $P^*$ is the distribution function with bounded point masses and support points given by $(p_i^*,x_i^*)_{i \notin \mathcal I}$, is a feasible solution of Program \eqref{opt: lambda infty finite support}. Using a similar argument as in the derivation of inequality \eqref{ineq : bounded support const} and the fact that $H$ is upper semi-continuous in Assumption \ref{assump:semi-cont H}, we can show that $Z^* \leq \sum_{i \notin \mathcal I} p_i^{*}H\left(x_i^{*}\right)  + \lambda_{M} s^* $. As a result,  $Z^* < \infty$ when $H$ is bounded on any compact subset of $\mathbb R^+$, and  \eqref{opt: lambda infty finite support}  returns an upper bound to $Z^*$ since $(P^*, s^*)$ is a feasible solution of \eqref{opt: lambda infty finite support}.

We now prove that $Z^*$ is also an upper bound to \eqref{opt: lambda infty finite support}.  We do so by noting that the LHS in \eqref{ineq: implicit const J} is equal to $\sum_{j \in \mathcal J \setminus \mathcal J(\infty)}  y_j \lambda_{jM} $. Hence, the implicit constraint $\sum_{j \in \mathcal J \setminus \mathcal J(\infty)}  y_j \lambda_{j,M}  \geq \lambda_M$ must hold for \eqref{opt : SILP dual drop Jx Jinfty} and we can add it to the constraint set of the latter. Based on Theorem \ref{theorem:strong duality} and Remark \ref{rmk: strong duality}, strong duality also holds for \eqref{opt : SILP dual drop Jx Jinfty} when the constraint $\sum_{j \in \mathcal J \setminus \mathcal J(\infty)}  y_j \lambda_{j,M} \geq \lambda_M $ is added to the formulation.  We then have
 \begin{equation}
 \begin{array}{*{5}l}
Z^* =\underset{P,s}{\sup}&\int H dP + \lambda_M s\\
\text{subject to}& \int G_j dP  + \lambda_{j,M}s \leq \gamma_j\text{\ \ for all $j \in \mathcal J \setminus \mathcal J(\infty)$}\\
& P \in \mathcal Q(\mathbb R^+), s\geq 0
\end{array}\label{opt:include s}
\end{equation}

The feasible region of \eqref{opt:include s} includes that of \eqref{opt: lambda infty finite support}, so \eqref{opt: lambda infty finite support} is bounded above by $Z^*$. Hence, \eqref{opt : SILP} and \eqref{opt: lambda infty finite support} have the same optimal objective value. This concludes our proof.\hfill \Halmos\\
\endproof

\begin{remark}\label{rmk: strong duality}
Under Assumption \ref{slater}, strong duality continues to hold for \eqref{opt : SILP new} when $x = \infty$ and \eqref{opt:include s}, i.e. the optimal value of
\begin{equation}
\begin{array}{llll}
&\underset{P}{\sup}&\int H dP \\
&\text{subject to\ \ }& \int G_{j} dP \leq \gamma_{j} &\text{\ \ for all $j \in \mathcal J \setminus \mathcal J(\infty)$}\\
&& P \in \mathcal Q(\mathbb R^+)
&\end{array}\label{primal1}
\end{equation}
is equal to that of
$$\begin{array}{lllll}
\underset{y}{\inf} &\sum_{j \in \mathcal J \setminus \mathcal J(\infty)}  y_j \gamma_j  \\
\text{subject to}&\sum_{j \in \mathcal J \setminus \mathcal J(\infty)} y_j G_j(u) &\geq H(u)&\text{\ \ for all\ }u\in \mathbb R^+\\
&y_j &\geq 0 & \text{\ \ for all\ } j \in \mathcal J \setminus \mathcal J(\infty)
\end{array}$$
and the optimal value of
\begin{equation}
\begin{array}{*{5}l}
\underset{P,s}{\sup}&\int H dP + \lambda_M s\\
\text{subject to}& \int G_j dP  + \lambda_{j,M}s \leq \gamma_j\text{\ \ for all $j \in \mathcal J\setminus\mathcal J(\infty)$} \\
& P \in \mathcal Q(\mathbb R^+), s\geq 0
\end{array}\label{primal2}
\end{equation}
is equal to that of
$$\begin{array}{lllll}
Z^* = \underset{y}{\inf} &\sum_{j \in \mathcal J \setminus \mathcal J(\infty)}  y_j \gamma_j  \\
\text{subject to}&\sum_{j \in \mathcal J \setminus \mathcal J(\infty)} y_j G_j(u) &\geq H(u)&\text{\ \ for all\ }u\in \mathbb R^+\\
& \sum_{j \in \mathcal J \setminus \mathcal J(\infty)}  y_j \lambda_{j,M} &\geq \lambda_M \\
&y_j &\geq 0 & \text{\ \ for all\ } j \in \mathcal J \setminus \mathcal J(\infty)
\end{array}$$
To see these, note that \eqref{primal1} can be similarly written in the form \eqref{strong duality form} but with less inequalities than those in $\tilde{\mathcal J}$ and some equalities in $\tilde{\mathcal J}'$ becoming inequalities. The interior point conditions there can be verified to hold for these new reduced set of constraints. On the other hand, \eqref{primal2} has the same form as \eqref{primal1} except that we can view the decision variable (e.g., the distribution) as having support on $\mathbb R^+$ together with a point mass $s$ on one augmented point. The interior point conditions held for \eqref{primal1} can be translated to this case by merely considering $s=0$.
\end{remark}

 \begin{theorem}\label{theorem:discrete  support}
Let  $H:\mathbb R^+ \to \mathbb R$ and $G_j:\mathbb R^+ \to \mathbb R$ be measurable functions for all $j$ in a finite index set $\mathcal J$. Then programs \eqref{opt : SILP bounded measure} and \eqref{opt : SILP bounded measure discrete support} below have the same optimal objective value:
\begin{equation}
 \begin{array}{lll}
\underset{P}{\sup}&\int H dP \\
\text{subject to\ \ } &\int dP \leq \nu\\
& \int G_{j} dP \leq \gamma_{j} &\text{\ \ for all $j \in \mathcal J$}\\
& P \in \mathcal Q(\mathbb R^+)
\end{array}\label{opt : SILP bounded measure}
\end{equation}

\begin{equation}
 \begin{array}{lll}
\underset{P}{\sup}&\int H dP \\
\text{subject to\ \ } & \int dP \leq \nu\\
& \int G_{j} dP \leq \gamma_{j} &\text{\ \ for all $j \in \mathcal J$}\\
& P \in \mathcal Q_{N}(\mathbb R^+)
\end{array}\label{opt : SILP bounded measure  discrete support}
\end{equation}
where $\nu \in \mathbb R^+$,  $\gamma_j \in \mathbb R$ for all $j \in \mathcal J$, and
 $N$ is the number of linearly independent functions in the sequence $\{(G_j)_{j\in \mathcal J}, 1\}$.
\end{theorem}

\proof{}
We  partition the feasible region of \eqref{opt : SILP bounded measure} into two subregions, one with the additional constraint $\int dP>0$, and another with $\int dP=0$. We consider two programs, each one the same as  \eqref{opt : SILP bounded measure} but with the respective additional constraint. Clearly, the maximum of these two programs have the same optimal value as \eqref{opt : SILP bounded measure}. We show that each program is equivalent to \eqref{opt : SILP bounded measure  discrete support} with the corresponding additional constraint, and since the maximum of these equivalent programs has the same optimal value as \eqref{opt : SILP bounded measure  discrete support}, we conclude the theorem.

In the case $\int dP=0$, the equivalence trivially holds. In the alternate case $\int dP > 0$,  the inequality $\int dP \leq \nu$ becomes  $\int dP = \nu - s$ for some $s \in [0,\nu)$. By applying a change of variable $P(x) := P(x)/(\nu-s)$,  the subprogram considered here can  be reformulated as
\begin{equation}
 \begin{array}{llll}
\underset{s \in [0,\nu)}{\sup}&\underset{P}{\sup}& (\nu -s)\int H dP \\
&\text{subject to\ \ } &\int dP = 1\\
&& \int G_{j} dP \leq \gamma_{j}/{(\nu-s)} &\text{\ \ for all $j \in \mathcal J$}\\
&& P \in \mathcal Q(\mathbb R^+)
\end{array}\label{opt : SILP bounded measure reformulated}
\end{equation}

The feasible region of the inner  program in \eqref{opt : SILP bounded measure reformulated}  is the set of probability measures defined on $\mathbb R^+$. Since all probability measures on in $\mathbb R^+$ (a Polish space) are regular, Theorem \ref{theorem:winkler adapted} applies to conclude that  \eqref{opt : SILP bounded measure reformulated} is equivalent to
\begin{equation}
 \begin{array}{llll}
\underset{s \in [0,\nu)}{\sup}&\underset{P}{\sup}&(\nu-s)\int H dP \\
&\text{subject to\ \ } &\int dP = 1\\
&& \int G_{j} dP \leq \gamma_{j}/{(\nu-s)} &\text{\ \ for all $j \in \mathcal J$}\\
&& P \in \mathcal Q_{N}(\mathbb R^+)
\end{array}\label{opt : SILP bounded measure reformulated discrete}
\end{equation}
By changing back the variable, we see that  \eqref{opt : SILP bounded measure reformulated discrete} is the same as \eqref{opt : SILP bounded measure  discrete support} with the additional constraint  $\int dP > 0$. We therefore conclude our theorem.\hfill \Halmos
\endproof

\begin{theorem}[A particular case of Theorem 3.2 \cite{winkler1988extreme}]
Let $\mathcal X$ be a Hausdorff space, $\mathcal F$ be the Borel $\sigma$-field,  $P_r(\mathcal X)$ be the set of regular probability measures on $\mathcal X$. In addition, let $f_1, \ldots, f_n$ be measurable functions, $c_1,\ldots, c_n$ are real values,  and
$$\mathcal H=\left\{q\in P_r(\mathcal X):f_i\text{\ is $q$-integrable and\ }\int f_idq\leq c_i,\ 1\leq i\leq n\right\}$$
In addition, let $g$ be a function on $\mathcal X$ integrable for every $q\in\mathcal H$ (possibly with integral values $\infty$ or $-\infty$). Then,
$$\sup\left\lbrace\int_{\mathcal X} gdq :q\in\mathcal H\right\rbrace=\sup\left\lbrace\int_{\mathcal X} gdq :q\in\text{ex\ }\mathcal H\right\rbrace$$
where $\text{ex\ }\mathcal H$ denotes the set of all extreme points of $\mathcal H$, i.e.
\begin{eqnarray*}
\text{ex\ }\mathcal H&=\Bigg\{&q\in\mathcal H:q=\sum_{i=1}^N t_i\cdot\delta(x_i),\ t_i>0,\ \sum_{i=1}^N t_i=1,\ x_i\in\mathcal X,\ 1\leq N \leq n+1,\\
&&\text{the vectors\ }(f_1(x_i),\ldots,f_n(x_i),1),\ 1\leq i\leq N,\text{\ are linearly independent}\Bigg\}
\end{eqnarray*}

\label{theorem:winkler adapted}
\end{theorem}
\proof{}
By Proposition 3.1 \cite{winkler1988extreme}, $G(q) = \int_{\mathcal X} g dq$ is a measure affine functional and Theorem 3.2 of \cite{winkler1988extreme} holds. In addition, Examples 2.1(a) in \cite{winkler1988extreme} mentions that the set $P$ in Theorem 2.1 of \cite{winkler1988extreme} can be chosen to be the set of all regular probability measures. As such, the extreme points of $\mathcal H$ in  Theorem 3.2 of \cite{winkler1988extreme} are precisely the ones defined in Theorem 2.1(a) of \cite{winkler1988extreme}.\hfill \Halmos\\
\endproof

\begin{remark}\label{rmk: sufficient theorem Pk}
In the proof of Theorem \ref{theorem:existence bounded measure}, Assumption \ref{assump:postivity} is only used to ensure the boundedness of the sequence $P^{(k')}$. In fact, Theorem \ref{theorem:existence bounded measure} would still hold provided that $\liminf_{k\to \infty}\int dP^{(k')} < \infty$. In this case, there would be a subsequence $k''$ such that $\int dP^{(k'')} < \infty$ and the rest of the proof would remain valid. 
\end{remark}

\proof{Proof of Corollary \ref{corly:moment LP}:}
Program  \eqref{opt : SILP not condensed} can be reformulated as
\begin{equation}
\begin{array}{lll}
\underset{P}{\sup} &\int H dP \\
\text{subject to\ \ }&  \int G_{j,1}  dP \leq \overline \gamma_{j,1} &\text{\ \ for all } j \in \mathcal J_1 \\
& \int -G_{j,1}  dP \leq -\underline \gamma_{j,1} &\text{\ \ for all } j \in \mathcal J_1 \\
&  \int G_{j,2}  dP \leq \overline \gamma_{j,2} &\text{\ \ for all } j \in \mathcal J_2 \\
& \int -G_{j,2}  dP \leq -\underline \gamma_{j,2} &\text{\ \ for all } j \in \mathcal J_2 \\
&P \in \mathcal Q(\mathbb R^+)\\
\end{array}\label{opt : SILP momemts organized}
\end{equation}
We verify all the assumptions needed to invoke Theorem \ref{theorem:existence bounded measure}. By definition, $G_{j,1}$ and $G_{j,2}$ are polynomials for all $j$, and by our choice of $M$, $\lim_{x\to\infty} G_{j,1}(x)/|G_{M,1}(x)| = \delta_{j,M}$ is well-defined and finite for all $j \in \mathcal J_1$. The same holds for  $\lim_{x\to\infty} G_{j,2}(x)/|G_{M,1}(x)|$ which is null for all $j\in \mathcal J_2$. Thus Assumptions \ref{assump:regularityinfty} and \ref{assump:well defined} hold. Since $\limsup_{x\to\infty}  G_{M,1}(x)/|G_{M,1}(x)| = 1 > 0$, we also have that the set $\mathcal J(\infty)$ is empty, so that cases 1 and 2 in Theorem \ref{theorem:existence bounded measure} does not occur. In addition, Assumption \ref{assump:semi-cont G} is trivially verified.

When $D \geq 1$, the function $H(x) = x^{J-D}(h \circ u_a)^{(-D)}(x)$ is continuous for $x>0$. From the generalized L'H\^{o}spital's rule, we can also show that $H$ is bounded on any compact subset of $\mathbb R^+$ since
\begin{equation}\label{ineq: hospital}
\liminf_{x\to 0} (h \circ u_a)^{(-J)}(x) \leq \liminf_{x\to 0} H(x) \leq \limsup_{x\to 0} H(x) \leq \limsup_{x\to 0} (h \circ u_a)^{(-J)}(x)
\end{equation}
Assumption \ref{assump:semi-cont H}  therefore holds in this case since both ends of \eqref{ineq: hospital} are bounded by the definition of $(h \circ u_a)^{(-J)}$. In particular, they are equal to $0$ when $J \geq 1$. When $D = 0$, Assumption \ref{assump:semi-cont H}  is also satisfied since we have assumed $h$ upper semi-continuous in this case.

Lastly, Assumption \ref{assump:postivity} is also satisfied since $G_{J,2} = 1/(D-J)!$ when $\mathcal J_2$ is not empty and $G_{0,1} = 1/D!$ otherwise, which correspondingly can serve as the constraint function needed in Assumption \ref{assump:postivity}.

From Theorem \ref{theorem:existence bounded measure} item 3a, we have $Z^* = \infty$ if $\lambda_M = \limsup_{u\to\infty} H(u)/|G_{M,1}(u)| = \infty$.  Otherwise, Theorem \ref{theorem:existence bounded measure} item 3b concludes that  program \eqref{opt : SILP momemts organized} can be reformulated as
\begin{equation}
\begin{array}{lll}
\underset{P, s}{\sup} &\int H dP  +\lambda_M s \\
\text{subject to\ \ }&  \int G_{j,1}  dP+ s \delta_{j M} \leq \overline \gamma_{j,1} &\text{\ \ for all } j \in \mathcal J_1 \\
& \int -G_{j,1}  dP  - s \delta_{j M}\leq -\underline \gamma_{j,1} &\text{\ \ for all } j \in \mathcal J_1 \\
& \int G_{j,2}  dP \leq \overline \gamma_{j,2} &\text{\ \ for all } j \in \mathcal J_2 \\
& \int -G_{j,2}  dP \leq -\underline \gamma_{j,2} &\text{\ \ for all } j \in \mathcal J_2 \\
& s = 0 \text{ if } \lambda_M = -\infty\\
& s \geq 0\\
&P \in \mathcal Q_{N}(\mathbb R^+)\\
\end{array}\label{opt : SILP momemts organized lambda_M lambda_m}
\end{equation}
which is equivalent to \eqref{opt:finite LP moments no J_i empty}.\hfill \Halmos\\ 
 \endproof

\proof{Proof of Corollary \ref{corly:existence P*}:} We prove the corollary by showing that \eqref{opt : SILP} admits, in either case, a sequence of feasible solutions $P^{(k)}$ such that $\int |G|dP^{(k)} \to \infty$ and $\int H dP^{(k)}\to Z^*$.

\noindent \textit{Item 1:} In this case, \eqref{opt : SILP} has a sequence of feasible solution  $P^{(k)}$  such that $\int H dP^{(k)} \to Z^*= \infty$.  If $\liminf_{u\to\infty} |G(u)/H(u)| = l$ where $l > 0$, then for all $\varepsilon \in (0,l)$  and $x$ large enough,
\begin{equation}
\int_x^{\infty} |G| dP^{(k)} = \int_x^{\infty} \left|\frac{G}{H}\right| |H| dP^{(k)} \geq \int_x^{\infty}\left|\frac{G}{H}\right| H dP^{(k)} \geq (l-\varepsilon)\int_x^{\infty} H dP^{(k)}
\end{equation}
which converges to $\infty$ as $k$ grows, and so $\int |G| dP^{(k)} \to \infty$.

\noindent \textit{Item 2:} We start by proving that the following program is unbounded.
\begin{equation}
  \begin{array}{lll}
 \underset{P}{\sup}&\int |G|dP\\
 \text{subject to\ \ }& \int G_j dP \leq \gamma_{j} &\text{\ \ for all $j \in \mathcal J$}\\
 & P \in \mathcal Q(\mathbb R^+)
 \end{array}\label{opt : existence P* Z* infinite}
 \end{equation}
We do so by applying Theorem \ref{theorem:existence bounded measure} with the function $H$ set to $|G|$. If $\mathcal J = \mathcal J(\infty)$,  we are in the case of Theorem \eqref{theorem:existence bounded measure}\eqref{theorem:item empty positive} since $G$ is not identically $0$ and therefore $\sup_{x\in \mathbb R^+} |G(x)| > 0$. Otherwise,  we are in Theorem \ref{theorem:existence bounded measure}\eqref{theorem:item lambdam inf lambda M inf} since we have $\limsup_{x\to\infty}|G(x)/G_M(x)| = \infty$. In either case, we obtain that \eqref{opt : existence P* Z* infinite} is unbounded. So there exists a sequence of feasible solution  $P^{(k)}$ for  \eqref{opt : SILP} satisfying $\int |G|dP^{(k)} \to \infty$. In addition, we have $\limsup_{k\to\infty} \int dP^{(k)} < \infty$. To see this, note that for all $j \in \mathcal J$
\begin{equation}
\inf_{u\in \mathbb R^+} G_j(u) \int dP^{(k)} \leq \int G_j dP^{(k)} \leq \gamma_j
\end{equation}
By assumption, there exists $j \in \mathcal J$ such that $\inf_{x\in \mathbb R^+} G_j(x) > 0$. As a result,
\begin{equation}
\limsup_{k\to\infty} \int_0^x |G| dP^{(k)} \leq \sup_{u \in [0,x]} |G(u)|\limsup_{k\to\infty} P^{(k)}(x) \leq \gamma_j\sup_{u \in [0,x]} |G(u)| /\inf_{u\in \mathbb R^+} G_j(u) < \infty
\end{equation}
for all $x \in \mathbb R^+$. Consequently, $\lim_{k \to\infty} \int_x^\infty |G|dP^{(k)} = \infty$ for all $x \in \mathbb R^+$. Furthermore, if $\liminf_{u\to\infty} H(u)/|G(u)| = l > 0$, then  for all $\varepsilon \in (0,l)$ and $x$ large enough,
\begin{align}
\int H dP^{(k)} &= \int_0^x H dP^{(k)} + \int_x^\infty H dP^{(k)} \nonumber\\
						&\geq \inf_{u \in [0,x]} H(u)P^{(k)}(x) + \int_x^\infty \frac{H}{|G|} |G| dP^{(k)}\nonumber\\
						& \geq \inf_{u \in [0,x]} H(u)P^{(k)}(x) + (l-\varepsilon)\int_x^\infty  |G| dP^{(k)} \label{ineq: corly existence P*}
\end{align}
The first and second terms in the RHS are respectively finite and unbounded when $k$ grows. So the LHS goes to $\infty$ with $k$.

 \noindent \textit{Item 3:} Using a similar argument as the proof of Item 2, we apply Theorem \ref{theorem:existence bounded measure}\eqref{theorem:item empty positive}  and  \ref{theorem:existence bounded measure}\eqref{theorem:item lambdam inf lambda M inf} to show that
\begin{equation}
  \begin{array}{lll}
 \underset{P}{\sup}&\int |G|dP\\
 \text{subject to\ \ }& \int G_j dP \leq \gamma_{j} &\text{\ \ for all $j \in \mathcal J$}\\
 &\int H dP =  Z^*\\
 & P \in \mathcal Q(\mathbb R^+)
 \end{array}\label{opt : existence P* Z* finite}
 \end{equation}
is unbounded, where $Z^*$ is the optimal objective value of \eqref{opt : SILP}. This implies the existence of a sequence of distribution functions  $P^{(k)}$ satisfying $\lim_{k\to\infty}\int |G|dP^{(k)} = \infty$, $\int G_j dP^{(k)} \leq \gamma_j$ for all $j \in \mathcal J$, and $\int H dP^{(k)} \to Z^*$. Because \eqref{opt : infinite constraint}  is bounded above by \eqref{opt : SILP}, this concludes that \eqref{opt : infinite constraint}  and \eqref{opt : SILP} have the same optimal objective value.\hfill \Halmos
 \endproof
 \newpage
\section{Phase I - Obtaining a Feasible Solution for Program \eqref{opt: lambda infty finite support}}\label{sec : algo Phase I}
\begin{algorithm}[H]
\caption{Finding a feasible solution of \eqref{opt: lambda infty finite support} when $\mathcal J \setminus \mathcal J(\infty)$ is not empty}
  \label{GLP initialization}
  \textbf{Inputs:} Provide the parameters $\gamma_j$ and the functions $G_j$ for all $j \in \mathcal J\setminus \mathcal J(\infty)$, and the function $H$. Compute the quantities $\lambda_{j,M},j \in \mathcal J\setminus \mathcal J(\infty)$ and $ \lambda_M$. Also specify a big number $C \in  \mathbb R^+$.\\

   \textbf{Initialization: }
\begin{itemize}
\item SET $x_1$ to an arbitrary value of the set $[0,C]$
\end{itemize}

  \textbf{Procedure: }For each iteration $k=1,2,\ldots$, given $(x_i)_{i = 1\ldots k}$:

  \begin{algorithmic}
  \State \textbf{1. }(Master problem) Solve
  $$\begin{array}{llll}
Z^k = &\underset{ s, r, p}{\min} & r\\
&\text{subject to}& - r + \sum_{i=1}^k G_j(x_i)p_i +  \lambda_{j,M} s\leq \gamma_j ,& \forall j \in \mathcal J\setminus \mathcal J(\infty)\\
&&s \geq 0\\
&& s = 0 \text{ if } \lambda_M  = -\infty\\
&&r, p_i\geq 0  & \forall i=1,\ldots,k
\end{array}$$
Let $((p_i^{k})_{i = 1, \ldots, k}, r^{k},  s^{k})$ be the optimal solution. Find the dual multipliers  $(y_j^k)_{j \in \mathcal J\setminus \mathcal J(\infty)}$ of dual multipliers  satisfying
$$\begin{array}{lll}
&\left(\sum_{j \in \mathcal J\setminus \mathcal J(\infty)} y_j^k G_j(x_i)\right)p_i^{k} = 0,& \ \forall i=1,\ldots,k\\
&\left(\sum_{j \in \mathcal J\setminus \mathcal J(\infty)} y_j^k\lambda_{j,M}\right)s^{k} = 0\\
&\left(- 1 - \sum_{j \in \mathcal J\setminus \mathcal J(\infty)}y_j^k \right)r^{k}  = 0\\
&y_j^k \geq 0 &\text{for all $j\in \mathcal J \setminus \mathcal J(\infty)$}
\end{array}$$

\State \textbf{2. }(Subproblem) Find $x_{k+1}$  that minimizes
$$\rho^k(u)=  \sum_{j \in \mathcal J\setminus \mathcal J(\infty)} y_j^k G_j(u), \ \ \ \text{where } u \in [0,C]$$

\begin{itemize}
\item SET $\epsilon^k = \rho^k(x_{k+1})$
\item IF $\epsilon^k  \leq 0$ and $r = 0$, STOP and RETURN $(x_i, p_i,^{(k)} s^{(k)})_{i=1,\ldots, k}$
\item IF $\epsilon^k  \leq 0$ and $r > 0$, STOP; The problem is inconsistent.
\end{itemize}\\

  \end{algorithmic}

\end{algorithm}

Note that if we focus on program \eqref{opt : SILP} instead of \eqref{opt: lambda infty finite support}, we can similarly run Algorithm \ref{GLP initialization}, but setting $\lambda_{j,M}=0$ for all $j \in \mathcal J \setminus \mathcal J(\infty)$, to obtain a feasible solution. However, as we have discussed,  \eqref{opt : SILP} operates on an unbounded domain and may not bear an optimal solution to whom the algorithm can converge.

\end{document}